\documentclass[iop,floatfix]{emulateapj}

\RequirePackage{color}
\RequirePackage{code}
\input{astro.sty}

\usepackage[T1]{fontenc}

\shorttitle{PS1 Data Processing System}
\shortauthors{E.A. Magnier et al}
\begin{document}
\title{The Pan-STARRS Data Processing System}

\def\IfA{1}
\def\LSST{2}
\def\Princeton{3}
\def\DUR{4}
\def\CfA{5}

\author{
Eugene A. Magnier,\altaffilmark{\IfA}
K.~C. Chambers,\altaffilmark{\IfA} 
H.~A. Flewelling,\altaffilmark{\IfA}
J.~C. Hoblitt,\altaffilmark{\LSST}
M. E. Huber,\altaffilmark{\IfA}
P.~A. Price,\altaffilmark{\Princeton}
W.~E. Sweeney,\altaffilmark{\IfA}
C. Z. Waters,\altaffilmark{\IfA}
L. Denneau,\altaffilmark{\IfA}
P. Draper,\altaffilmark{\DUR}
K. W. Hodapp,\altaffilmark{\IfA}
R. Jedicke,\altaffilmark{\IfA}
N. Kaiser,\altaffilmark{\IfA}
R.-P. Kudritzki,\altaffilmark{\IfA}
N. Metcalfe,\altaffilmark{\DUR}
C.~W. Stubbs,\altaffilmark{\CfA}
R. J. Wainscoat\altaffilmark{\IfA}
} 

\altaffiltext{\IfA}{Institute for Astronomy, University of Hawaii, 2680 Woodlawn Drive, Honolulu HI 96822}
\altaffiltext{\LSST}{LSST Project Management Office, Tucson, AZ, U.S.A}
\altaffiltext{\Princeton}{Department of Astrophysical Sciences, Princeton University, Princeton, NJ 08544, USA}
\altaffiltext{\DUR}{Department of Physics, Durham University, South Road, Durham DH1 3LE, UK}
\altaffiltext{\CfA}{Harvard-Smithsonian Center for Astrophysics, 60 Garden Street, Cambridge, MA 02138}
\begin{abstract}

The Pan-STARRS Data Processing System is responsible for the steps
needed to downloaded, archive, and process all images obtained by the
Pan-STARRS telescopes, including real-time detection of transient
sources such as supernovae and moving objects including potentially
hazardous asteroids.  With a nightly data volume of up to 4 terabytes
and an archive of over 4 petabytes of raw imagery, Pan-STARRS is
solidly in the realm of Big Data astronomy.  The full data processing
system consists of several subsystems covering the wide range of
necessary capabilities.  This article describes the Image Processing
Pipeline and its connections to both the summit data systems and the
outward-facing systems downstream.  The latter include the Moving
Object Processing System (MOPS) \& the public database: the Published
Science Products Subsystem (PSPS).

\end{abstract}

\keywords{Surveys:\PSONE }

\section{Introduction}
\label{sec:intro}

The 1.8m Pan-STARRS\,1 telescope is located on the summit of Haleakala
on the Hawaiian island of Maui.  The wide-field optical design of the
telescope \citep{2004SPIE.5489..667H} produces a 3.3 degree field of view with
low distortion and minimal vignetting even at the edges of the
illuminated region.  The optics and natural seeing combine to yield
good image quality: 75\% of the images have full-width half-max values
less than (1.51, 1.39, 1.34, 1.27, 1.21) arcseconds for (\grizy), with
a floor of $\sim 0.7$ arcseconds.

The \PSONE\ camera \citep{2009amos.confE..40T}, known as GPC1, consists of a
mosaic of 60 back-illuminated CCDs manufactured by Lincoln Laboratory.
The CCDs each consist of an $8\times8$ grid of $590 \times 598$
pixel readout regions, yielding an effective $4846 \times 4868$
detector.  Initial performance assessments are presented in
\cite{2008SPIE.7014E..0DO}.  Routine observations are conducted remotely from the
Advanced Technology Research Center in Kula, the main facility of the
University of Hawaii's Institute for Astronomy (IfA) operations on Maui.
The Pan-STARRS1 filters and photometric system have already been
described in detail in \cite{2012ApJ...750...99T}.

For nearly 4 years, from 2010 May through 2014 March, this telescope
was used to perform a collection of astronomical surveys under the
aegis of the Pan-STARRS Science Consortium.  The majority of the time
(56\%) was spent on surveying the $\frac{3}{4}$ of the sky north of
$-30$ Declination with \grizy\ filters in the so-called $3\pi$ Survey.
Another $\sim 25\%$ of the time was concentrated on repeated deep
observations of 10 specific fields in the Medium-Deep Survey.  The
rest of the time was used for several other surveys, including a
search for potentially hazardous asteroids in our solar system.  The
details of the telescope, surveys, and resulting science publications
are described by \cite{chambers2017}.

Pan-STARRS produced its first large-scale public data release, Data
Release 1 (DR1) on 16 December 2016.  DR1 contains the results of the
third full reduction of the Pan-STARRS $3\pi$ Survey archival data,
identified as PV3.  Previous reductions (PV0, PV1, PV2) were used
internally for pipeline optimization and the development of the
initial photometric and astrometric reference catalog
\citep{magnier2017.calibration}.  The products from these reductions
were not publicly released, but have been used to produce a wide range
of scientific papers from the Pan-STARRS 1 Science Consortium members
\citep{chambers2017}.  DR1 contained only average information
resulting from the many individual images obtained by the $3\pi$
Survey observations.  A second data release, DR2, was made available
28 January 2019.  DR2 provides measurements from all of the individual
exposures, and include an improved calibration of the PV3 processing
of that dataset.

This is the second in a series of seven papers describing the
Pan-STARRS1 Surveys, the data reduction techniques and the resulting
data products.  This paper (Paper II) presents a description of the
Pan-STARRS data handling systems, with an emphasis on the Image
Processing Pipeline (IPP).  The Pan-STARRS Image Processing Pipeline
consists of a suite of software programs and data systems that are
designed to reduce astronomical images, measure astronomical sources
on the images, perform the calibration, and distribute the results to
various users.  The processing system includes extensive
parallelization across a large cluster of computers in order to
process the large amount of data generated by the Pan-STARRS\,1
telescope.


\citet[][Paper I]{chambers2017}
provide an overview of the Pan-STARRS System, the design and
execution of the Surveys, the resulting image and catalog data
products, a discussion of the overall data quality and basic
characteristics, and a brief summary of important results.


\citet[][Paper III]{waters2017} describe the details of the pixel
processing algorithms, including detrending, warping, and adding (to
create stacked images) and subtracting (to create difference images)
and resulting image products and their properties.

\citet[][Paper IV]{magnier2017.analysis} describe the details of the
source detection and photometry, including point-spread-function and
extended source fitting models, and the techniques for ``forced''
photometry measurements.

\citet[][Paper V]{magnier2017.calibration}
describe the final calibration process, and the resulting photometric and astrometric quality.  

\citet[][Paper VI]{flewelling2017}
describe the details of the resulting catalog data and its
organization in the Pan-STARRS database.

\citet[][Paper VII]{huber2017} describe the Medium Deep Survey in
detail, including the unique issues and data products specific to that
survey. The Medium Deep Survey is not part of Data Releases 1 or 2 and
will be made available in a future data release.

Section~\ref{sec:overview} provides an overview of the full data
analysis system and breaks down the major elements of the Image
Processing Pipeline.  Section~\ref{sec:stages} discusses in some
detail each of the analysis steps which may be applied to the images
and resulting catalogs of detected sources.
Section~\ref{sec:postprocessing} discusses the databasing system used
for calibration, the calibration operations, and summarizes the
construction of the public release database.
Section~\ref{sec:operations} discusses the operational infrastructure
of the IPP.  Section~\ref{sec:hardware} discusses the hardware systems
used by the IPP for regular nightly operations and for processing the
PV3 data release, with some details on the scale of computing needed
to reduce this large number of exposures.  



\section{Overview of Pan-STARRS Data Processing}
\label{sec:overview}

\subsection{Elements of the Pan-STARRS Data Processing System}

The Pan-STARRS data analysis system consists of many elements to
support a wide range of activities: archiving and management of the
raw and processed image files; real-time nightly processing of images
for transient and moving object science; large-scale re-processing and
calibration to produce measurements for the science collaboration and
the wider public; specialized image processing to facilitate research
and development of the analysis system itself; and distribution of the
resulting data products to various consumers in a variety of formats
and modes.

The Pan-STARRS data analysis system is divided internally into several major
components:
\begin{itemize}
\item Summit Processing : both the camera and observatory summit systems perform
  data analysis tasks needed to support the on-going observations.
  In this article, we focus only on those aspects used by the off-summit
  analysis stages.
\item Image Processing Pipeline (IPP) : this portion of the data
  analysis system takes the data from raw pixels on the summit
  computers to calibrated measurements of astronomical objects in an
  internal databasing system.
\item Moving Object Processing System (MOPS) : this system is
  responsible for linking individual detections of solar-system
  objects together and determining the orbits \citep[][]{2013PASP..125..357D}.
\item Published Science Products Subsystem (PSPS) : this system
  ingests the calibrated measurements from the IPP, MOPS, and others
  and generates a high-availability database with web-based
  interactions for public consumption \citep[][]{flewelling2017}.

\end{itemize}
Management of the above set of analysis stages takes place at the IfA
within the scope of responsibility of the Pan-STARRS Observatory.
Across the wider Pan-STARRS collaboration(s), additional data analysis
operations are performed to support science results.  These
collaboration-wide analysis operations range from those which are
tightly coupled to the Pan-STARRS Observatory system, such as the
analysis of the transient search teams and the public archive database
at MAST, to those which perform offline analysis for eventual ingest
back into the Pan-STARRS databases and archive.  The latter category
includes the ubercal photometric analysis \citep{2012ApJ...756..158S}, the photo-z
analysis \citep{2012ApJ...746..128S}, and the QSO / RR Lyra search efforts
\citep{2016ApJ...817...73H}.  In addition, collaborations within the wider
Pan-STARRS community have implemented a variety of science-level
analyses of their own to support their science goals \citep[e.g., M31
variable search][]{2014ApJ...797...22L,2012AJ....143...89L}.

\begin{figure*}[htbp]
  \begin{center}
 \includegraphics[width=\hsize,clip]{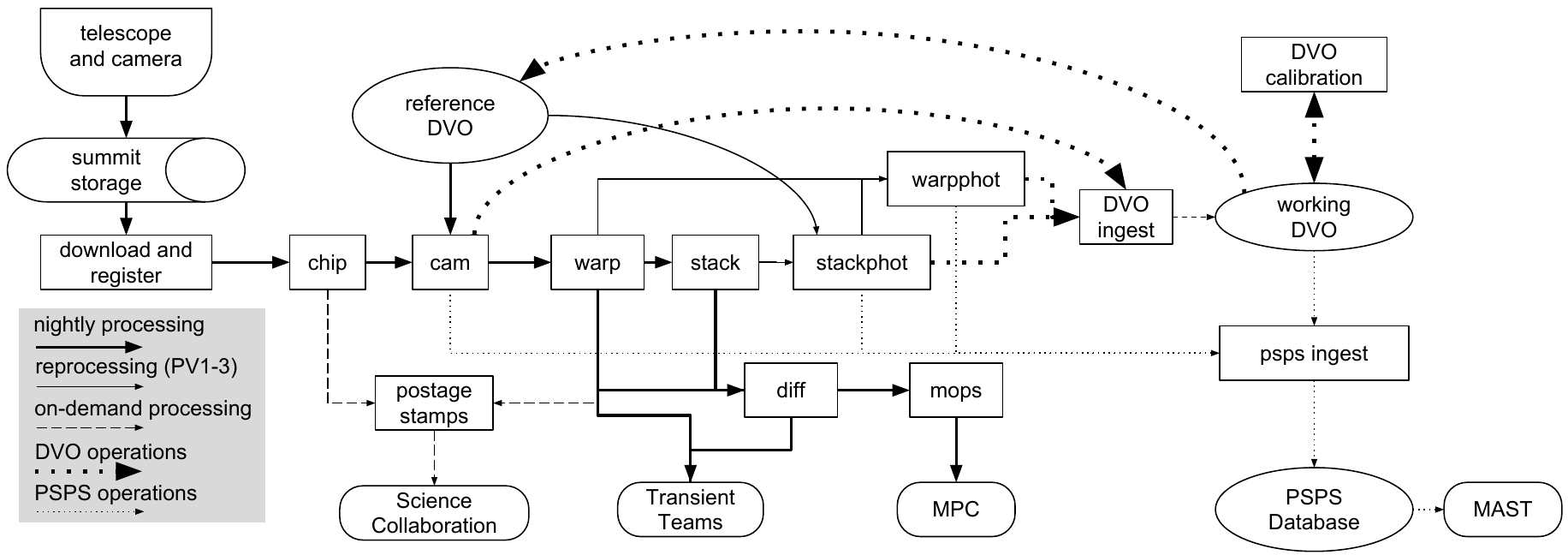}
  \caption{\label{fig:analysis.elements} Elements of the Pan-STARRS\,1
    Data Analysis System.  Rectangles represent data analysis steps;
    ellipses represent databases; rounded rectangles represent
    external groups (``customers'').  The arrows show a simplified representation
  of the major flow of data between the analysis stages and data
  processing elements.}
  \end{center}
\end{figure*}

Figure~\ref{fig:analysis.elements} illustrates the many elements of
the Pan-STARRS data analysis system.  This figure focuses on the data
analysis steps which occur within the Pan-STARRS Observatory, with an
emphasis on the analysis, calibration, and database ingest stages.
The MOPS is described in detail by \cite{2013PASP..125..357D}.


\subsection{Nightly Processing Analysis Stages}

Data analysis to support nightly science operations is driven by two
main goals: 1) rapid detection of the moving and transient sources to
enable recovery or follow-up with other telescopes. 2) regular
analysis of the images to monitor data quality and for use in
longer-timescale science projects.  Not all of the analysis elements
listed in Figure~\ref{fig:analysis.elements} are used by the nightly
analysis system.  Each of the data analysis stages are discussed in
detail below.  In short, each image is processed independently to
correct for instrumental signatures and to detect the astronomical
sources (\IPPstage{chip}); astrometric and photometric calibrations
are determined (\IPPstage{camera}), and finally images are
geometrically transformed to a common pixel representation
(\IPPstage{warp}).  Warped images may either be added together
(\IPPstage{stack}) or used in an image subtraction (\IPPstage{diff}).
As part of nightly science processing, images for certain fields such
as the Medium Deep survey fields \citep[see][]{huber2017}, are stacked
together in nightly chunks, providing deeper detection capability on
1-day timescales.  Depending on the survey mode, difference images are
generated for the nightly stack images (using a deep stack template)
or for individual warp images.  In the later case, the warp images may
be differenced against another warp from the same night or against a
reference stack from the appropriate part of the sky.

\subsection{Re-processing Analysis Stages}

Pan-STARRS has performed several large-scale reprocessings of both the
Medium Deep and $3\pi$ Survey data for internal consumption.  For the
$3\pi$ Survey data, we identify these large-scale reprocessings as PV1
(Processing Version 1), PV2, and PV3, with PV3 the analysis used for
the first public data release, DR1.  We also refer to the nightly
science analysis of the data as PV0.  For these reprocessing stages,
the standard steps of \ippstage{chip} through \ippstage{warp}, plus
\ippstage{stack} and \ippstage{diff} are performed, starting from raw
data, usually using a single homogeneous version of the data analysis
procedures.  PV2 was a special case in which we started from the
camera level products of PV1 to speed up the turn-around to the
community.  In addition to the analysis stages listed above which are
shared with the nightly processing, these large-scale reprocessing
analyses include additional processing steps.  A more detailed
photometric analysis is performed on the stacks, including
morphological analysis appropriate to galaxies (model fits, Kron and
Petrosian aperture photometry, etc).  The results of the stack
photometry analysis are used to drive a forced-photometry analysis of
the warp images.  These analysis steps are discussed in detail by
\citet[][]{magnier2017.analysis}.  The data products from the camera,
stack, and forced-warp photometry analysis stages are ingested into
the internal calibration database (DVO, the Desktop Virtual
Observatory) and used for photometric and astrometric calibrations
\citet[see Section~\ref{sec:DVO} and][]{magnier2017.calibration}.

\subsection{Data Access and Distribution}

During the PS1 Science Consortium operations, data products were
provided to the consortium members from many different stages of the
analysis process.  Data access by the PS1 Science Consortium members
was managed through a variety of mechanisms depending on the data
volume and type of data products desired.
Figure~\ref{fig:analysis.elements} illustrates some of these
connections.  Access to small samples of imaging data was provided on
demand via the Postage Stamp server; access to large sets of
pre-defined raw and reduced data products was provided via the
Distribution and Publication systems.  The internal calibration DVO
databases were provided at several stages via a separate DVO
distribution mechanism.  For the first two large-scale reprocessings
(PV1 \& PV2), the data were ingested into the PSPS database system and
made available to the PS1SC community through a web portal based at
the IfA as well as the MAST portal \citep[see][for full
  details]{flewelling2017}.

\section{IPP Data Processing Stages}
\label{sec:stages}

\subsection{Processing Database}
\label{sec:processing.database}

\begin{table*}
\caption{GPC1 Database Schema Outline}
\begin{center}
\footnotesize
\begin{tabular}{llll}
\hline
\hline
{\bf Stage} & {\bf Primary Table} & {\bf Secondary Table(s)} & {\bf Key} \\
  \ippstage{summitcopy}   & \ippdbtable{pzDataStore}  &                                  & \\
                          & \ippdbtable{summitExp}    & \ippdbtable{summitImfile}        & \ippdbcolumn{summit_id} \\
                          & \ippdbtable{pzDownloadExp}& \ippdbtable{pzDownloadImfile}    & \\
                          & \ippdbtable{newExp}       & \ippdbtable{newImfile}           & \ippdbcolumn{exp_id} \\

  \ippstage{registration} & \ippdbtable{rawExp}       & \ippdbtable{rawImfile}           & \ippdbcolumn{exp_id} \\
  \ippstage{chip}         & \ippdbtable{chipRun}      & \ippdbtable{chipProcessedImfile} & \ippdbcolumn{chip_id} \\
  \ippstage{camera}       & \ippdbtable{camRun}       & \ippdbtable{camProcessedExp}     & \ippdbcolumn{cam_id} \\
  \ippstage{fake}         & \ippdbtable{fakeRun}      & \ippdbtable{fakeProcessedImfile} & \ippdbcolumn{fake_id} \\
  \ippstage{warp}         & \ippdbtable{warpRun}      & \ippdbtable{warpImfile}          & \ippdbcolumn{warp_id} \\
                          &                           & \ippdbtable{warpSkyCellMap}      & \\
                          &                           & \ippdbtable{warpSkyfile}         & \\
  \ippstage{stack}        & \ippdbtable{stackRun}     & \ippdbtable{stackInputSkyfile}   & \ippdbcolumn{stack_id} \\
                          &                           & \ippdbtable{stackSumSkyfile}     & \\
  \ippstage{staticsky}    & \ippdbtable{staticskyRun} & \ippdbtable{staticskyInput}      & \ippdbcolumn{sky_id} \\
                          &                           & \ippdbtable{staticskyResult}     & \\
  \ippstage{skycal}       & \ippdbtable{skycalRun}    & \ippdbtable{skycalResult}        & \ippdbcolumn{skycal_id} \\
  \ippstage{fullforce}    & \ippdbtable{fullForceRun} & \ippdbtable{fullForceInput}      & \ippdbcolumn{ff_id} \\
                          &                           & \ippdbtable{fullForceResult}     & \\
                          &                           & \ippdbtable{fullForceSummary}    & \\
  \ippstage{diff}         & \ippdbtable{diffRun}      & \ippdbtable{diffSkyfile}         & \ippdbcolumn{diff_id} \\
                          &                           & \ippdbtable{diffInputSkyfile}    & \\
  \ippstage{detrend}      & \ippdbtable{detRun}       & \ippdbtable{detRunSummary}       & \ippdbcolumn{det_id} \\
                          &                           & \ippdbtable{detInputExp}         & \\
                          &                           & \ippdbtable{detRegisteredImfile} & \\
                          &                           & \ippdbtable{detStackedImfile}    & \\
                          & \ippdbtable{detProcessedExp} & \ippdbtable{detProcessedImfile}  & \\
                          & \ippdbtable{detResidExp}  & \ippdbtable{detResidImfile}      & \\
                          & \ippdbtable{detNormalizedExp} & \ippdbtable{detNormalizedImfile} & \\
  \ippstage{addstar}      & \ippdbtable{addRun}       & \ippdbtable{addProcessedExp}     & \ippdbcolumn{add_id} \\
  \ippstage{distribution} & \ippdbtable{distRun}      & \ippdbtable{distComponent}       & \ippdbcolumn{dist_id} \\
                          &                           & \ippdbtable{distTarget}          & \\
  \ippstage{publish}      & \ippdbtable{publishRun}   & \ippdbtable{publishDone}         & \ippdbcolumn{pub_id} \\
                          &                           & \ippdbtable{publishClient}       & \\
  \ippstage{lap}          & \ippdbtable{lapSequence}  & \ippdbtable{lapRun}              & \ippdbcolumn{seq_id} \\
                          & \ippdbtable{lapRun}       & \ippdbtable{lapExp}              & \ippdbcolumn{lap_id} \\
  \ippstage{remote}       & \ippdbtable{remoteRun}    & \ippdbtable{remoteComponent}     & \ippdbcolumn{remote_id} \\
\hline
\end{tabular}
\label{tab:database_schema}
\end{center}
\end{table*} 

A critical element in the Pan-STARRS IPP infrastructure is the
processing database.  This database, using the mysql database engine,
tracks information about each of the processing stages.  It is used as
the point of mediation of all IPP operations.  Processing stages
within the IPP perform queries of the database to identify the data to
be processed at a given stage.  As the processing for a particular
stage is completed, summary information about the stage is written
back to the database.  In this way, the database records this history
of the processing, and also provides the information needed by
successive processing stages to begin their own tasks.

The processing database is colloquially referred to as the ``gpc1''
database, since a single instance of the database is used to track the
processing of images and data products related to the PS1 GPC1 camera.
This same database engine also has instances (same schema, different
data) for other cameras processed by the IPP, e.g., GPC2, the test
cameras TC1, TC3, and the Imaging Sky Probe (ISP).  In general,
processing information for different cameras is separate in different
processing database; merging of output products takes place in DVO.

Within the processing database, the various processing stages are
represented as a set of tables.  In general, there is a top level
primary table which defines the conceptual list of processing items
either to be done, in progress, or completed.  An associated secondary
table (or set of tables) lists the details of component elements which
have been processed for each top-level item.  Table
\ref{tab:database_schema} contains an outline of the database schema,
showing the relations between tables organized by processing stage.
As an example, one critical stage is the \ippstage{chip} processing
stage (see \S\ref{sec:chip}) in which the individual chips from an
exposure are detrended and sources are detected.  Within the gpc1
database, the primary table is called \ippdbtable{chipRun} in which
each exposure has a single entry.  Associated with this table is the
\ippdbtable{chipProcessedImfile} table, which contains one row for
each of the chips associated with the exposure (up to 60 for gpc1).
The primary tables, such as \ippdbtable{chipRun}, are populated once
the system has decided that a specific item (e.g., an exposure) should
be processed at that stage.  Initially, the entry is given a state of
``run'', denoting that the exposure is ready to be processed.  The
low-level table entries, such as the \ippdbtable{chipProcessedImfile}
entries, are only populated once the element (e.g., the chip) has been
processed by the analysis system.  Once all elements for a given
stage, e.g., chips in this case, are completed, then the status of the
top-level table entry (\ippdbtable{chipRun}) are switched from ``run''
to ``full''.

If the analysis of an element (e.g., the individual OTA chip)
completed successfully, then the corresponding table row (e.g.,
\ippdbtable{chipProcessedImfile}) is listed with a \ippdbcolumn{fault}
of 0.  If the analysis failed, then a non-zero \ippdbcolumn{fault} is
recorded.  An analysis which results in a \ippdbcolumn{fault} is one
in which the failure is thought to be temporary.  For example, if the
processing computer had a network interruption and was unable to write
some of the output files, this would be an ephemeral failure which was
not a failing of the data, but merely the processing system.  On the
other hand, if the analysis failed because of a problem with the input
data, this is noted by setting a non-zero value in a different table
field, \ippdbcolumn{quality}.  For example, if the \ippstage{chip} analysis
failed to discover any stars because the image was completely
saturated, the analysis can complete successfully (\ippdbcolumn{fault}
= 0), but the \ippdbcolumn{quality} field will be set to a non-zero
value.  The various processing stages are able to select only the good
(\ippdbcolumn{quality} = 0) elements of a prior stage when choosing
items for processing.  For example, the \ippstage{camera} calibration
stage will only use data from chips with good \ippdbcolumn{quality}
data, dropping the failed chips from the rest of the analysis.  On the
other hand, a \ippdbcolumn{fault} in one of the elements for a given
stage will block any successive stages which depend on that result
from processing that item.  In this way, if such a temporary failure
occurs, the system will not process an exposure through subsequent
stages without the component that has failed temporarily.  Since many
of the \ippdbcolumn{fault}s which occur are ephemeral due to current
conditions of the processing cluster, the processing stages are set up
to occasionally clear and re-try the faulted entries.  Some \ippdbcolumn{fault}s
represent software bugs and in the early stages of processing were
accumulated until the corresponding software issue could be addressed;
since the start of the PS1 Science Consortium Surveys, these types of
\ippdbcolumn{fault}s have largely been eliminated.  Thus, automatic processing is
able to keep the data flowing even in the face of occasional network
glitches or hardware crashes.

\subsection{Summit copy}
\label{sec:summitcopy}

As exposures are taken by the PS1 telescope \& GPC1 camera system, the
data from the 60 OTA devices are read out by the camera software
system and written to disk on a collection of computers at the summit
in the PS1 facility called ``pixel servers.'' After the images are
written to disk, a summary listing of the information about the
exposure and the chip images are added to the summit datastore (an
internal http-based data sharing tool, see
Section~\ref{sec:datastore}).

During night-time operations, while the summit datastore is being
populated, the IPP subsystem called \ippstage{summitcopy} monitors the
datastores listed in the \ippdbtable{pzDatastore} table of the
database in order to discover new exposures ready for download.  Once
a new exposure has been listed on the datastore, \ippstage{summitcopy}
adds an entry of the exposure to a table in the processing database
(\ippdbtable{summitExp}), indexed by an identifier that simply
increments the number of exposures announced by the summit, the
\ippdbcolumn{summit_id}.  This tells the \ippstage{summitcopy} system
to look for the list of chips, which are then added to another table
(\ippdbtable{summitImfile}).  This system then attempts to download
the chips (registering the results of those operations into the
\ippdbtable{pzDownloadExp} and \ippdbtable{pzDownloadImfile} tables)
from the summit pixel servers via an http request.  As the image files
are downloaded, their MD5 checksum values are calculated and compared
with the value reported by the summit datastore.  Download failures
are rare and marked with a non-zero \ippdbcolumn{fault}, allowing for
a manual recovery, rather than automatically rejecting the failed
chips.  Once all the components of the exposure have been downloaded,
they are further entered into the \ippdbtable{newExp} and
\ippdbtable{newImfile} tables, which index the exposures by
\ippdbcolumn{exp_id}.  This switch in index indicates that the
exposure has successfully been copied from the summit to the IPP
cluster, and that further processing is no longer dependent on outside
resources.

\subsection{Image Registration}
\label{sec:registration}

Once the chips for an exposure have all been downloaded, the exposure
is ready to be registered.  In this context, ``registration'' refers to
the process of adding them to the database listing of known, raw
exposures (not to be confused with ``registration'' in the sense of
pixel re-alignment).  The result of the \ippstage{registration} analysis is an
entry for each exposure in the \ippdbtable{rawExp} table, and one for
each chip in the \ippdbtable{rawImfile} table.  These tables are
critical for downstream processing to identify what exposures are
available for processing in any other stage.  At the \ippstage{registration}
stage, a large amount of descriptive metadata for each chip is added
to the \ippdbtable{rawImfile} table, the majority of which is
extracted from the chip FITS file headers (e.g., RA, DEC, FILTER) and
some of which is determined by a quick analysis of the pixels (e.g.,
mean pixel values, standard deviation).  The chip-level information is
merged into a set of exposure-level metadata and added to the
\ippdbtable{rawExp} table entry.  The exposure-level metadata may be
the same as any one of the chip, in a case where the values are
duplicated across the chip files (e.g., the name of the telescope or
the date \& time of the exposure), or it may be a calculation based on
the values from each chip (e.g., average of the average pixel values).

Unlike much of the rest of the IPP stage, the raw exposures may only
have a single entry in the \ippstage{registration} tables of the processing
database tables (\ippdbtable{rawExp} and \ippdbtable{rawImfile}).

For GPC1, the \ippstage{registration} stage is also the stage at which
the \ippprog{burntool} analysis is run.  This analysis is more
completely described in \citet{waters2017}.  In brief, the
\ippprog{burntool} program identifies bright sources on the image, and
identifies persistence trails that result from the incomplete transfer
of charge.  As this charge can leak out in subsequent exposures, the
burntool analysis is run sequentially on the exposures, based on the
observation date and time listed in the headers, with the results
stored on disk.  As a result of the sequential nature of this
analysis, the \ippstage{registration} of exposures is blocked until
the \ippprog{burntool} has been run on the previous exposures.

Once the \ippstage{registration} process has finished, new science
exposures that have an \ippdbcolumn{obs_mode} value that indicates
they are part of a particular science survey are automatically
launched into the science analysis by defining entries for the
\ippstage{chip} processing stage, as described above.  The science
analysis of a given exposure can be relaunched multiple times, such as
for the large scale PV3 reprocessing.  The automatically-launched
analysis process ensures the shortest time between observation and
analysis, particularly important in the search for transient sources.

\subsection{Chip Processing}
\label{sec:chip}

The science analysis of an exposure begins with the \ippstage{chip}
stage, which operates on the individual OTA image files.  This
analysis step has two main goals: detrending the image to remove the
instrumental signature from the pixel values, and the detection of
astronomical sources in the objects.  Based on the entry the
\ippdbtable{chipRun} primary table defining the processing details
(with the \ippdbcolumn{state} column indicating it needs processing),
and the associated information listed in the \ippdbtable{rawImfile},
jobs can be spawned for each component OTA.  

The \ippstage{chip} stage is naturally parallelized by processing data
from each of the 60 OTAs independently.  Several stages in the IPP
analysis are parallelized in a similar fashion; although there are
multiple stages that operate on an entire exposure at once, the
majority of stages operate on smaller segments of a full exposure,
allowing the processing tasks to be spread over the machines in the
processing cluster.  The \ippprog{pantasks} environment (the system
which manages the processing jobs, see Section~\ref{sec:pantasks})
attempts to target the processing to a computer which is assigned to
host data for the particular OTA.  This capability is implemented to
reduce the network I/O load by minimizing the number of operations
done on non-local data.  In practice, this targeted processing has not
had as large of an impact as was originally intended: the data volume
and operational details of the hardware has reduced the ability of any
one node to reliably contain a particular OTA.  The targeted
processing has probably reduced the network load somewhat but it has
not been as critical of a requirement as originally expected.


The actual image processing is performed by the \ippprog{ppImage}
program.  This program reads the raw data into memory and applies the
detrend corrections \citep[see][]{waters2017} to each cell in the OTA
(stored as different extensions in the FITS file format), and then
mosaics the cells into a single contiguous \ippstage{chip} stage
image.  This step also creates in memory additional images to hold the
mask data, which indicates which pixels may not be valid, and the
variance image, constructed as the Poissonian noise on the number of
electrons detected based on the original pixel value and the detector
gain.  A background model is then fit across the image and subtracted
to remove the expected contribution from the sky
\citep[see][]{waters2017} for details.

With the image calibration procedure finished, object identification
and photometry can be performed.  Although this can be done using a
stand alone program, \ippprog{psphot}, the underlying functions are
contained in a library that allows \ippprog{ppImage} to directly do
this analysis, removing the need to write out and re-read the image
data.  The details of the detection and characterization of the
sources in the image are provided in \citet{magnier2017.analysis}.  

The results of the image processing are then written to disk,
including the science, mask, and variance images, the binned
background model subtracted, the PSF model used in the photometry
process, and a FITS catalog of detected sources.  Additional binned
images of the full OTA are also saved, using $16\times{}16$ and
$256\times{}256$ pixel binning scales for quick visualization.  The
processing log and a selection of summary metadata describing the
processing results are also written to disk.  This metadata is used to
populate a row in the \ippdbtable{chipProcessedImfile} table to
indicate that the processing of this OTA is complete.

As each OTA is processed independently of the others across a number
of computers, the \ippprog{pantasks} server managing the jobs
periodically runs an \ippmisc{advance} task that checks that the
number of rows in \ippdbtable{chipProcessedImfile} with
\ippdbcolumn{fault} equal to zero matches the associated number of
rows in \ippdbtable{rawImfile}.  If this condition is met, than all
processing for that exposure is finished, and the \ippdbcolumn{state}
field is set to ``full''.  If the
\ippdbtable{chipRun}.\ippdbcolumn{end_stage} field is set to
\ippstage{chip}, then no further action is taken.  However, this field
is usually set to a subsequent stage (most often \ippstage{warp}),
in which case an entry for this exposure is added to the \ippdbtable{camRun}
table, and processing continues.



\subsection{Camera Calibration}
\label{sec:camera}

After sources have been detected and measured for each of the chips,
the next stage is to perform a basic calibration of the full exposure
in the \ippstage{camera} stage.  This runs as a single job for the
entire exposure, passing the collection of FITS table catalogs
generated from each OTA in the \ippstage{chip} stage to the
\ippprog{psastro} program.  Although the full catalog is loaded, the
calibration primarily concerns the positions ($x_{\rm ccd}, y_{\rm
  ccd}$) and the instrumental PSF magnitudes.  The header information
in these catalogs is used to determine the coordinates of the
telescope boresite (RA, DEC, position angle).  These three coordinates
are used, along with a pre-determined model of the OTA layout within
the camera, to generate an initial guess for the astrometry of each
chip.  Reference star coordinates and magnitudes are loaded from a
reference catalog for a region corresponding to the boundaries of the
exposure, padded by a large fraction (25\%) of the exposure diameter
to help guarantee a solution in the case of a modest pointing error.
The guess astrometry is used to match the reference catalog to the
observed stellar positions in the focal plane coordinate system
\citep[see][]{magnier2017.calibration}.  Early on in the PS1SC
mission, the nightly processing (PV0) used a reference catalog based
on a combination of external catalogs (2MASS, Tycho, USNO).  Later, 
reference catalogs based on Pan-STARRS data was used.  For the $3\pi$ PV3 analysis,
the reference catalog was based on Pan-STARRS data from the PV2
analysis \citep[see][for more details]{magnier2017.calibration}.

Once an acceptable match is found, the astrometric calibration of the
individual chips is performed, including a fit to a single model for
the distortion introduced by the camera optics.  The astrometric model
includes a set of 3rd order polynomials for the transformations from the chip
coordinate system to the camera focal plane coordinate system and a
single additional 3rd order polynomial transformation from the camera focal
plane coordinate system to the tangent plane of a tangent projection.
For the $3\pi$ PV3 analysis, the typical astrometric residuals are in
the range of 20 - 30 milliarcseconds, sufficient to match observations
of the same objects between different exposures.  There are, however,
inevitable outliers.  Certain chips occasionally have systematically worse
astrometry, with OTA XY17 notably poor in this respect.

After the astrometic analysis is completed, the photometric
calibration is determined using the final match to the reference
catalog.  A single photometric zero point is determined for each
exposure, with the airmass term fixed to the nominal linear slope for
each filter.  No color terms are measured between the observed
photometry and the reference photometry.  However, at this stage,
pre-determined color terms may be used to transform the reference
photometry to an appropriate photometric system.  For the PS1 nightly
processing, the reference catalog does not include \wps\ photometry,
so a fixed color transformation is used to generate synthetic w-band
photometry from the \rps\ \& \ips\ photometry.  For more details, see
\cite{magnier2017.calibration}.  The result of these calibrations is
stored as a single multi-extension FITS table containing the results
from each OTA as a separate extension.

In addition to the astrometric and photometric calibrations, the
\ippstage{camera} stage also generates the dynamic masks for the
images.  These include masking for optical ghosts, glints, saturated
stars, diffraction spikes, and electronic crosstalk.  The mask images
generated by the \ippstage{chip} stage are updated with these dynamic
masks and a new set of files are saved for the downstream analysis
stages.  The \ippstage{camera} stage also merges the binned chip
images (see~\ref{sec:chip}) into single jpeg images of the entire
focal plane.  These jpeg images can then be displayed by the process
monitoring system to visualize the data processing.

Again, summary metadata is saved to disk as well, and the results
listed therein are used to populate a row in the
\ippdbtable{camProcessedExp} database table.  As the full exposure is
processed all at once, this update also updates the associated
\ippdbtable{camRun} entry, linked by the \ippdbcolumn{cam_id}.  As
with the \ippstage{chip} stage, if the
\ippdbtable{camRun}.\ippdbcolumn{end_stage} is for a subsequent
stage, an appropriate entry is added to the \ippdbtable{fakeRun}
table.  

\subsection{Fake Analysis}
\label{sec:fake}

The \ippstage{fake} stage was originally designed to do false source
injection and recovery, in order to determine the detection efficiency
of sources on the exposure.  However, early in the design of the IPP,
this task was moved to the rest of the photometry analysis done at the
\ippstage{chip} stage.  Removing the stage would require significant
changes to the database schema.  As a result, this conveniently named
stage generally does no actual data processing, and consists mainly of
database operations to move the exposure on to the \ippstage{warp}
stage.  The operations mimic the \ippstage{chip} stage, with
individual jobs run for each OTA that update rows in the
\ippdbtable{fakeProcessedImfile}, and an \ippmisc{advance} task that
updates the \ippdbtable{fakeRun} table and promotes the exposure to
the next stage by adding a row to the \ippdbtable{warpRun} table.

\subsection{Image Warping}
\label{sec:warp}

The \ippstage{warp} stage transforms the image pixels from the regular
grid laid out on the chips in the camera to a system of pixels with
consistent geometry for a location on the sky.  The new image
coordinate system is defined by one of a number of ``tessellations''
which specify how the sky is divided into individual images.  A single
tessellation starts with a collection of projection centers
distributed across the sky.  A grid of image pixels about each
projection center corresponds to sky positions via a projection with a
specified pixel scale and rotation.  In general, the pixel grid within
the projection is defined as a simplified grid with the y-axis aligned
to the Declination lines and no distortion terms.  The projection
centers are typically separated by several degrees on the sky; for
pixel scales appropriate to GPC1, the resulting collection of pixels
would be unwieldy in terms of memory in the processing computers.  The
pixel grid is thus subdivided into smaller sub-images called
`skycells'.

A tessellation can be defined for a limited region, with only a small
number of projection centers (e.g., for processing the M31 region), or
even a single projection center (e.g., for the Medium Deep fields).
For the $3\pi$ survey, the tessellation contains projection centers
covering the entire sky.  The version used for the PV3 analysis is
called the \ippmisc{RINGS.V3}.  In this tessellation, projection
centers are spaced every four degrees in DEC and the RA spacing is
approximately four degrees as well, adjusted to ensure an integer
number of equal-sized regions.  \ippmisc{RINGS.V3} uses a pixel scale
of $0\farcs{}25$ per pixel.  The projections are subdivided into a
$10\times{}10$ grid of skycells, with an overlap region of
60\arcsec\ between adjacent skycells to ensure that objects of modest
size are not split on all images.  The coordinate system used for
these images matches the parity of the sky, with north in the positive
$y$ direction and east to the negative $x$ direction.  The
tessellations used by the IPP are stored in the DVO format (see
Section~\ref{sec:DVO}).  A table in the processing database,
\ippdbtable{SkyTable}, lists the projection centers and image
boundaries for all skycells.

The first step of the \ippstage{warp} stage is to determine which
skycells overlap with the input exposure.  These overlaps are
determined by the \ippprog{dvoImageOverlaps} program, which compares
the astrometrically-calibrated catalog from the \ippstage{camera}
stage to the DVO database defining the target tessellation.  The
output of this command is used to populate the
\ippdbtable{warpSkyCellMap} table in the database, which contains a
row for each skycell and OTA that overlap.  Each skycell may contain
contributions from multiple OTAs; since they are similar in size, in a
typical situation the warp is constructed from 4-6 neighboring OTAs.

Once this mapping has been defined, jobs to warp the pixels onto each
skycell are run, passing the \ippstage{camera} stage catalog and the
\ippstage{chip} stage images (including the variance images and the
updated masks) to the \ippprog{pswarp} program.  For details on the
warping algorithm, see \cite{waters2017}.  The outputs of this program
are the geometrically transformed images (signal, variance, and mask)
containing all input pixels warped to the common skycell pixel grid,
These can subsequently be used for stacking and difference image
analyses.  For the $3\pi$ survey data, the signal, mask, and variance
images generated at this stage are being made available from the image
extraction tools at the MAST archive at STScI as part of the DR2 data
release.


When the \ippstage{warp} jobs have completed, an entry for the skycell
is added to the \ippdbtable{warpSkyfile} database table, linked to the
\ippdbtable{warpRun} entry by a common \ippdbcolumn{warp_id}.  An
\ippmisc{advance} task again checks that all potential skycells have
been generated.  At this point, the direct promotion of exposures from
one stage to the next stops, as the logic for matching exposures for
other combinations is more complicated than simply adding a single
entry.

\subsection{Stack Combination}
\label{sec:stack}

The skycell images generated by the \ippstage{warp} process can be
added together to make deeper, higher signal-to-noise images in the
\ippstage{stack} stage.  These stacked images also fill in coverage
gaps between different exposures, resulting in an image of the sky
with more uniform coverage than a single exposure.

In the IPP processing, stacks may be made with various options for the
input images.  During nightly science processing, the 8 exposures per
filter for each Medium Deep field are combined into a set of stacks
for that field.  These so-called ``nightly stacks'' are used by the
transient survey projects to detect faint supernovae, among other
transient events.  For the PV3 $3\pi$ analysis, all images in each
filter from the observations for this survey were stacked together to
generate a single set of images with $\sim 10 - 20\times$ the exposure
of the individual survey exposures.  

For the PV3 processing of the Medium Deep fields, stacks have been
generated for the nightly groups and for the full depth using all
exposures, producing ``deep stacks''.  In addition, a ``best seeing''
set of stacks have been produced using image quality cuts described by
\citet[][Paper VII]{huber2017}.  We have also generated out-of-season
stacks for the Medium Deep fields, in which all images {\em not} from a
particular observing season for a field are combined into a stack.
These later stacks are useful as deep templates when studying
long-term transient events in the Medium Deep fields as they are not
(or less) contaminated by the flux of the transients from a given
season.

When a given set of \ippstage{stack} stage processing is defined,
exposures with existing \ippstage{warp} entries that match the filter,
position, and other criteria such as seeing are identified.  An entry
is then added for each skycell in the \ippdbtable{stackRun} table,
with the \ippdbcolumn{warp_id} entries for the exposures added to the
\ippdbtable{stackInputSkyfile} table, linked to the
\ippdbtable{stackRun} entry by the \ippdbcolumn{stack_id} field.  This
defines the mapping for which exposures contribute to the
\ippstage{stack}.  The \ippstage{stack} stage processing is performed
at the skycell level.

The \ippstage{stack} jobs pass the information about the input images
and catalogs to the \ippprog{ppStack} program, which performs the
image combinations.  Input warps are combined based on a weighting
defined by the median variance for each image; see~\cite{waters2017}
for details on the stack combination algorithm.  In addition to the
standard image, mask, and variance produced at other stages,
additional images are constructed with information about the
contributions to each pixel.  A number image contains the number of
input exposures used for each pixel, along with an exposure time map,
and a weighted exposure time map that scales the exposure time based
on the relative variance of each input.  These images for the $3\pi$
analysis are currently available from the MAST image extraction tools
at STScI.

Upon completing the generation of these images, a row is added to the
\ippdbtable{stackSumSkyfile} table with statistics about
\ippstage{stack} processing.  As this completes all processing for the
entry, no \ippmisc{advance} job is required.

\subsection{Stack Photometry}
\label{sec:staticsky}

Although images are generated in the \ippstage{stack} stage of the
IPP, the source detection and analysis of those images is deferred to
the \ippstage{staticsky} stage.  This separation is maintained because
the photometry analysis of the \ippstage{stack} images, including
convolved galaxy model fitting, is performed on all 5 filters
simultaneously.  By deferring this analysis, the processing system may
also decouple the generation of the pixels from the source detection.
This makes the sequencing of analysis somewhat easier and less subject
to blocks due to a failure in the long-running stacking analysis.
Similar to the \ippstage{stack} stage, an entry is created in the
\ippdbtable{staticskyRun} table, linked to a series of rows in the
\ippdbtable{staticskyInput} table by a common \ippdbcolumn{sky_id},
each of which also contains the appropriate \ippdbcolumn{stack_id}
entries for the skycell under consideration.

The input images are passed to the \ippprog{psphotStack} program which
does the analysis.  The stack photometry algorithms are described in
detail in \cite{magnier2017.analysis}.  In short, sources are detected
in all 5 filter images down to the $5\sigma$ significance.  The
collection of detected sources is merged into a single master list.
If a source is detected in at least two bands, or only in \yps{} band,
then a PSF model is fitted to the pixels of the other bands in which
the source was not detected.  This forced photometry results in lower
significance measurements of the flux at the positions of objects
which are thought to be real sources, by virtue of triggering a
detection in at least two bands.  The relaxed limit for \yps{} band is
included to allow for searches of \yps{} dropout objects: it is known
that faint, high-redshift quasars may be detected in \yps{} band only.
Sources detected only in \yps{} band are therefore more likely to have
a higher false-positive rate than the other stack sources.  The
parameters of the PSF model are allowed to vary with position in the
skycell.  The PSF model is also used to convolve the analytical galaxy
models, which are the fitted to the observed flux distributions.
Galaxy models include S\'ersic, DeVaucouleur, and Exponential
profiles.

The stack photometry output files consist of a set of FITS table
catalogs, with one file for each filter.  Within these files, there
are multiple table extensions, with different classes of measurements
saved in the different extensions.  The extensions include a table of
the measurements of sources based on the PSF model; a table of
aperture-like parameters such as the Petrosian flux and radius; a
table of the convolved galaxy model fits; and a table of the radial
aperture measurements.  Once the photometry is complete, a row is
added to the \ippdbtable{staticskyResult} table with basic statistics
from the analysis.

The stack photometry output catalogs are re-calibrated for both
photometry and astrometry in a process very similar to the
\ippstage{camera} calibration stage.  Although the individual warps
which go into the stack are calibrated based on the \ippstage{camera}
stage analysis, there was some concern that these calibrations might
not be sufficiently well-defined for some of the input warps, biasing
the photometry of the stack.  By re-calibrating the stacks, we can be
sure that the stack photometry as measured is tied to the photometric
reference system.

In the case of this \ippstage{skycal} stage, each skycell is processed
independently.  Because of this independence, when queued for
processing, the entries in the \ippdbtable{skycalRun} table contain
the \ippdbcolumn{sky_id} and \ippdbcolumn{stack_id} entries of the
parent data directly.  As in the \ippstage{camera} stage, the
\ippprog{psastro} program reads in the stack photometry catalog, and
produces a calibrated output, with format matching the input.  A
different processing recipe is supplied to \ippprog{psastro}, which
controls for the different data.  The same reference catalog is used
for the \ippstage{camera} and \ippstage{stack} calibration stages.
Upon completion, the analysis statistics are written to the
\ippdbtable{skycalResult} table.

\subsection{Forced Warp Photometry}
\label{sec:fullforce}

Traditionally, projects which use multiple exposures to increase the
depth and sensitivity of the observations have generated something
equivalent to the \ippstage{stack} images produced by the IPP analysis
(c.f, CFHT Legacy survey, COSMOS, etc).  In theory, the photometry of
the \ippstage{stack} images produces the ``best'' photometry catalog,
with best sensitivity and the best data quality at all magnitudes.  In
practice, these images have some significant limitations due to the
difficulty of modelling the PSF variations.  This difficulty is
particularly severe for the Pan-STARRS $3\pi$ survey stacks due to the
combination of the substantial mask fraction of the individual input
exposures, the large intrinsic image quality variations within a
single exposure, and the wide range of image quality conditions under
which data were obtained and used to generate the $3\pi$ PV3 stacks.

For any specific stack, the point spread function at a particular
location is the result of the combination of the point spread
functions for those individual exposures which went into the stack at
that point.  Because of the high mask fraction, the exposures which
contributed to pixels at one location may be somewhat different just a
few tens of pixels away.  In the end, the \ippstage{stack} images have
a effective point spread function which is not just variable, but
changing significantly on small scales in a highly textured fashion.

Any measurement which relies on a good knowledge of the PSF at the
location of an object needs to determine the PSF variations present in
the \ippstage{stack} image, or the measurement will be somewhat
degraded.  The highly textured PSF variations make this a very
challenging problem: not only would such a PSF model need to be highly
fine-grained, there would likely not be enough stars in a given
\ippstage{stack} image to determine the model at the resolution
required.  The IPP photometry analysis code uses a PSF model with 2D
variations using a grid of at most $6\times 6$ samples per skycell, a
number reasonably well-matched to the density of stars at most
moderate Galactic latitudes for the PS1 3$\pi$ depths.  This scale is
far too large to track the fine-grained changes apparent in the stack
images.

Thus PSF photometry and convolved galaxy model analysis in the stack
are degraded by the PSF variations.  Aperture-like measurements are in
general not as affected by the PSF variations, as long as the aperture
in question is large compared to the FWHM of the PSF.


The IPP analysis solves this problem by using the sources
detected in the stack images and performing forced photometry on the
individual warp images used to generate the stack.  This
\ippstage{fullforce} analysis is performed on all warps for a single
skycell and filter as a single unit within the processing database,
while individual warps are processed individually in parallel as
separate processing jobs.  A separate PSF model is determine for each
of the warp images so that the combined measurement is reliable.

When processing is queued for this stage, an entry is added to the
\ippdbtable{fullForceRun} primary database table with a reference to
the corresponding stack and \ippdbcolumn{skycal_id} entry that is the
input source of detections to be measured.  The \ippdbcolumn{warp_id}
values for the input \ippstage{warp} stage images that contributed to
the \ippstage{stack} associated with that \ippdbcolumn{skycal_id} are
then added to the \ippdbtable{fullForceInput} table, linked to the
primary table by the \ippdbcolumn{ff_id} identifier.  The individual
jobs for each warp are then run, which passes the \ippstage{warp}
stage image products along with the \ippstage{skycal} catalog to the
\ippprog{psphotFullForce} program.


The convolved galaxy models are also re-measured on the
\ippstage{warp} images by the \ippstage{fullforce} stage analysis.  In
this analysis, the galaxy models determined by the
\ippstage{staticsky} photometry analysis are used to seed the analysis
in the individual \ippstage{warp} images.  The purpose of this
analysis is the same as the \ippstage{fullforce} PSF photometry: the
PSF of the \ippstage{stack} image is poorly determined due to the
masking and PSF variations in the inputs.  Without a good PSF model,
the PSF-convolved galaxy models are of limited accuracy.

Upon completion of the forced photometry, an entry is added to the
\ippdbtable{fullForceResult} table with the processing statistics for
that combination of \ippdbcolumn{ff_id} and \ippdbcolumn{warp_id}.
The individual warp measurements are combined together to produce an
average warp photometry value for each object within the context of
the DVO object database system, including re-calibration of each warp
based on the tie to the average photometry of the objects measured in
the \ippstage{camera} stage.

Once all of the entries in the \ippdbtable{fullForceInput} table have
finished, a summary operation is run to combine the galaxy photometry
analysis measurements into a single value.  The output catalogs listed
in the \ippdbtable{fullForceResult} table are passed to the
\ippprog{psphotFullForceSummary} to calculate the averages of the
individual warp measurements, weighted by the signal-to-noise of the
flux measurements.  When this analysis completes, an entry is added to
the \ippdbtable{fullForceSummary}, and the \ippdbtable{fullForceRun}
entry is marked as completed.


\subsection{Difference Images}
\label{sec:diff}

Two of the primary science drivers for the Pan-STARRS system are the
search for hazardous asteroids and the search for Type Ia supernovae to
measure the history of the expansion of the universe.  Both of these
projects require the discovery of faint, transient source in the
images.  For the hazardous asteroids, and solar system studies in
general, the sources are transient because they are moving between
observations; supernovae are stationary but transient in brightness.
In both cases, the discovery of these sources can be enhanced by
subtracting a static reference image from the image taken at a certain
epoch.  The quality of such a difference image can be enhanced by
convolving one or both of the images so that the PSFs in the two
images are matched \citep[e.g.,][]{1998ApJ...503..325A}.

In the \ippstage{diff} stage, the IPP generates difference images for
appropriately specified pairs of images.  It is possible for the
difference image to be generated from a pair of \ippstage{warp} stage
images, from a \ippstage{warp} and a \ippstage{stack} of some variety,
or from a pair of \ippstage{stack} stage images.  During the PS1
survey, pairs of exposures, called TTI pairs \citep[see Survey
  Strategy in][]{chambers2017}, were obtained for each pointing within
a $\approx$ 1 hour period in the same filter, and to the extent
possible with the same orientation and boresite position.  The
standard PS1 nightly processing generated difference images from the
resulting pairs of \ippstage{warp} images.  The nightly processing
generated \ippstage{stack} images for the Medium Deep fields, and
these were combined with a template reference \ippstage{stack} image
to generate ``stack-stack diffs'' each night they were observed.  For
the PV3 $3\pi$ processing, the entire collection of \ippstage{warp}
stage images for the survey were combined with images generated by the
\ippstage{stack} processing to generate ``warp-stack diffs'', for
eventual public released.

When a \ippstage{diff} processing is defined, an entry is added to the
\ippdbtable{diffRun} table, and the appropriate input images are added
to the \ippdbtable{diffInputSkyfile} table, with one entry for each
skycell that is covered by the images.  For a \ippstage{diff}
generated from two \ippstage{warp} stage products, the input images
have their \ippdbcolumn{warp_id} values recorded in the
\ippdbcolumn{warp1} and \ippdbcolumn{warp2} for each skycell that
overlaps.  If two \ippstage{stack} stages are to be used in the
difference, their \ippdbcolumn{stack_id} entries are recorded in the
\ippdbcolumn{stack1} and \ippdbcolumn{stack2} fields.  As each
\ippstage{stack} only covers a single skycell, the \ippstage{diff} is
usually defined indirectly, using other information from the
\ippdbtable{stackRun} table to select appropriate
\ippdbcolumn{stack_id} values.  Similarly, \ippstage{diff} processing
is defined for the mixed case by creating entries that populate one of
\ippdbcolumn{warp1} and \ippdbcolumn{stack1} and populating one of
\ippdbcolumn{warp2} and \ippdbcolumn{stack2}.  In all cases, the
minuend of the subtraction to be performed is the ``1'' entry, and the
subtrahend is the ``2'' entry.

Jobs are created based on the entries of
\ippdbtable{diffInputSkyfile}, with the appropriate images and
catalogs passed to the \ippprog{ppSub} program.  This does the
subtraction, as well as the photometry of any sources detected in the
\ippstage{diff} image.  Sources may be detected as a positive source
(flux in the minuend is higher than the subtrahend) or as a negative
source (flux in the subtrahend is higher).  The algorithm used for PSF
matching is described in \citet{waters2017}.  Upon completion of these
jobs, statistics about the processing are written to an entry in the
\ippdbtable{diffSkyfile} table.  An \ippmisc{advance} checks for the
completion of all of the components listed in
\ippdbtable{diffInputSkyfile}, and marks the \ippdbtable{diffRun}
entry as such.

\section{Post-Processing : Database Ingest and Calibration}
\label{sec:postprocessing}

\subsection{DVO}
\label{sec:DVO}

\subsubsection{Overview}

The Pan-STARRS IPP uses an internal database system, distinct from the
publicly visible database system, to determine the association
between multiple detections of the same astronomical object and as
part of the astrometric and photometric calibration process.  This
database system, called the ``Desktop Virtual Observatory'' (DVO) was
developed originally for the LONEOS project
\citep{1995DPS....27.0110B}, and used as part of the CFHT Elixir
system \citep{2004PASP..116..449M}.  The capabilities of this
databasing system have been somewhat expanded for the Pan-STARRS
context.

DVO tracks three main classes of information: 1) average properties of
astronomical objects; 2) measurements of those objects (from which the
average properties are derived); 3) properties of the images which
provided some or all of the measurements.  In addition, certain
metadata tables define general features of the database.
Table~\ref{tab:DVO_schema} lists the full collection of database
tables used by DVO.


In the most basic implementation, a collection of measurements for
detections from a set of images is loaded into DVO along with the
metadata describing the images.  The latter includes properties such
as the exposure time, airmass, filter, time \& date of the exposure,
etc.  Critically, the image metadata includes an astrometric
transformation relating the detection coordinate on the image to the
coordinate on the sky.  As the collection of measurements are loaded
into DVO, the software constructs astronomical objects based on those
detections.  If images overlap, multiple observations of the same
astronomical object are grouped together.  Thus, a single DVO database
will contain a one-to-many relationship between the images and the
measurements and a many-to-one relationship between the measurements
and the derived astronomical objects.

%


\subsubsection{DVO Schema}

\begin{table*}[hb]
\begin{center}
\caption{DVO Database Tables\label{tab:DVO_schema}}
\begin{tabular}{ll}
\hline
\hline
{\bf Table Name} & {\bf Description} \\
\hline
Images               & The images that have objects in the DB. \\
Average              & Astronomical objects including their astrometric properties. \\
SecFilt              & Average photometry of the objects in multiple filters (one filter per row) \\
Measure              & Detections of sources identified with an Object, potentially linked to an image. \\
StarPar              & Stellar parameters determined by the Harvard group \citep{2015ApJ...810...25G} \\
Lensing              & Lensing (KSB) parameters and fixed circular aperture photometry from the warps \\
LensObj              & Average lensing and fixed circular aperture photometry \\
Galphot              & Result of galaxy model fits (forced galaxy models) \\
SkyRegions           & spatial distribution of tables \\
Photcodes            & Transformations between different photometric systems \\
Hosts                & computers used to store the tables \\
\hline
\end{tabular}
\end{center}
\end{table*}

\paragraph{Photcodes}

DVO has a special metadata table called \ippdbtable{photcode} which
identifies the photometry filter systems.  Entries in this table are
used to identify the source of measurements and images.  Each row in
the \ippdbtable{photcode} table includes a \ippdbtable{photcode}
name, a unique numerical ID, and information about that photometry
system.  

There are 3 classes of photcodes defined within the DVO system.  One
class of photcodes define the filter systems for the average
photometry measurements; these are called \ippmisc{SEC} photcodes.  A
second class of photcode is associated with measurements from a
specific camera for which image metadata is available are called
\ippmisc{DEP} photcodes.  There are also those measurements which come
from external data sources for which DVO does not have any information
to determine a calibration (e.g., instrumental magnitudes and detector
coordinates).  These measurements are reference values and are
assigned \ippmisc{REF} photcodes.

The names for \ippmisc{SEC} photcodes are the names of filter systems,
such as $g,r,i$ or $J,H,K$.  For \ippmisc{DEP} and \ippmisc{REF}
photcodes, the names are constructed from the name of a camera or
telescope (e.g., GPC1 or 2MASS), the name (or short-hand name) of a
filter (e.g., \gps{}), and an identifier for the detector, if not
unique (e.g., XY01 for a GPC1 OTA).  

Additional information is associated with each photcode to define the
nominal zero point and airmass slope, as well as color trends to
transform a measurement in the specific photcode to a common system.
For example, a \ippmisc{DEP} photcode GPC1.g.X01 would have the
nominal zero point (24.563) and airmass term (0.147).  The database
elements allow for individual chips to have different color terms to
bring them to a common filter system.

DVO ingest methods are defined for several large-scale surveys for
which the published data represent average properties derived from
multiple measurements, and for which the measurement-to-image
relationship is not provided.  Ingests methods have been defined, for
example, for 2MASS, WISE, Gaia, USNO-B.  In each of these cases, the
astrometric and photometric measurements are stored in the
\ippdbtable{Measure} table, with the data source identified by the
photcode of the measurement.

\paragraph{Measurement Tables}

In most cases, the individual measurements of the astronomical objects
are carried in the table \ippdbtable{Measure}.  For measurements from
PS1 in the PV3 / DR1 or DR2 databases, this would consist of values
determined by \ippprog{psphot} for each \ippstage{chip},
\ippstage{warp}, or \ippstage{stack} stage image.  Measurements for
other cameras processed by the IPP may also be included similarly in a
DVO database.  Measurements from other sources, such as SDSS, 2MASS,
or WISE, can also be included in this table, distinguished by their
different photcodes.

The \ippdbtable{Measure} table includes the instrumental magnitudes
for the PSF, aperture, and Kron photometry; raw position
(\ippdbcolumn{Xccd}, \ippdbcolumn{Yccd}) and second moments
(\ippdbcolumn{Mxx}, \ippdbcolumn{Myy}, \ippdbcolumn{Mxy}), along with
shape parameters of the PSF model at the position of the object
(\ippdbcolumn{FWx}, \ippdbcolumn{FWy}, \ippdbcolumn{theta}).  Metadata
about the exposure that the measurement was derived from is also
included, such as the exposure time, the date \& time of the
observation, airmass, azimuth, and \ippdbcolumn{photcode} information
specifying the filter.  The \ippdbtable{Measure} table also carries
the calibration magnitude offsets ($M_{\rm cal}$ and $M_{\rm flat}$,
discussed below) and the astrometrically calibrated position.
Astrometric offsets for several systematic corrections discussed below
are also defined for each measurement.  Photometry from
\ippstage{chip}, \ippstage{warp}, and \ippstage{stack} are all placed
in the same table with photcodes distinguishing the source.  Since
stacks and forced warp fluxes may have non-significant values, the
table is somewhat de-normalized: it also carries both magnitudes as
well as instrumental flux values for the PSF, aperture, and Kron
photometry.  In this case, we have chosen to trade storage space for
computing time.

For the warp images, we also measure the weak lensing KSB parameters
related to the shear and smear tensors \citep{1995ApJ...449..460K}.
These measurements are stored in the \ippdbtable{Lensing} table,
along with the radial aperture fluxes for radii numbers 5, 6, \& 7
(respectively 3.0, 4.63, and 7.43 arcsec).  This table contains one
row for every warp image on which the object was measured. 

The \ippdbtable{Galphot} table stores the results of the forced galaxy
fitting analysis for each object that has been measured.  This table
contains one row per filter and model type (S\'ersic, Exponential, or
DeVaucouleur) if forced galaxy models have been calculate for the
object.

The \ippdbtable{Starpar} table carries measurements provided by the
Harvard team (Green, Schlafly, Finkbeiner) from the analysis of the
SED of objects in the PS1 $3\pi$ data, using the PV2 analysis version
\citep{2015ApJ...810...25G,2014ApJ...783..114G}.  In this work, the
goal was a 3D model of the dust in the Galaxy based on Pan-STARRS and
2MASS photometry.  As part of this analysis, the authors fit the SEDs
of all stellar sources (as determined by a cut based on the PSF -
aperture magnitudes) with stellar models including free parameters of
extinction, distance modulus, metallicity, and absolute r-band
magnitude.  While these photometric distance modulus measurements are
not extremely precise, they provide a constraint on the distance which
is used in our analysis of the astrometry
\citep[see][]{magnier2017.calibration}.



\paragraph{Object Tables}
\label{sec:object}

One of the main purposes of DVO is to define the relationship between
individual detections of an astronomical object and the definition of
that object.  New detections are generally added to the database in a
group associated with, for example, an image from the GPC1 camera.  As
new detections are loaded, they are compared to the objects already
stored in the database.  If an object is already found in the database
within the match radius, the new detection is assigned to that
object. If more than one object exists within the database, the
detection is associated with the closest object.  For most data
sources, a match radius of 1.0 arcsecond is used, but this may be
adjusted in special cases.

Two tables carry the most important information about the astronomical
objects in the database: \ippdbtable{Average} and
\ippdbtable{SecFilt}.  These two tables specify the main average
properties of the astronomical object.  The \ippdbtable{Average} table includes the
astrometric information ($\alpha, \delta, \mu_\alpha, \mu_\delta,
\pi$) and associated errors, data quality flags for each object, links
to the other tables, and a number of IDs, with one row for each
astronomical object.  
The \ippdbtable{SecFilt} table\footnote{The name \ippdbtable{SecFilt}
  is a bit of a historical misnomer: originally, DVO was designed for
  a monochromatic survey and data for a single photometric band was
  maintained in the Average table.  Later, DVO was adapted to a
  multifilter system and additional filters were added to the
  \ippdbtable{SecFilt} (Secondary Filter) table.  Eventually, the
  schema was normalized and all photometric data placed in
  \ippdbtable{SecFilt}, with the Primary filter concept being dropped,
  but the name has since stuck.} contains average photometric
information for a collection of filters.  A given DVO instance has a
specified set of filters for which average photometry is stored in the
\ippdbtable{SecFilt} table.  The number and choice of filters for the
\ippdbtable{SecFilt} may be modified by the database administrator
fairly easily, but the process of updating the database is somewhat
expensive (\approx 24 hours for the current PV3 instance).  Thus the
choice is semi-static for a given DVO instance.  In the case of the
PV3 DVO instance, 9 average bandpasses are defined: \gps{}, \rps{},
\ips{}, \zps{}, \yps{}, {\it J, H, K}, and \wps{}.  The
\ippdbtable{SecFilt} table contains one row for each filter for each
object, thus the PV3 DVO contains 9 times as many rows as the
\ippdbtable{Average} table.  The order of the table is fixed in
relation to the \ippdbtable{Average} table: row $i$ of
\ippdbtable{Average} defines the object with photometry contained in
rows $9i \rightarrow 9i + 8$ ($i$ is zero counting).

The values stored in the \ippdbtable{Lensing} table are used to
calculate average values for each of these types of measurements in
each filter.  The \ippdbtable{Lensobj} table stores the averaged KSB
and radial aperture photometry for each of the 5 filters \grizy.  This
table contains one entry per object per filter.  The table is not
generally stored unsorted as it is calculated after the full database
is populated.  The link from \ippdbtable{Average} to
\ippdbtable{Lensobj} is defined by the fields
\ippdbtable{Average}.\ippdbcolumn{offsetLensobj} and
\ippdbtable{Average}.\ippdbcolumn{Nlensobj}.  Each
\ippdbtable{Lensobj} row also includes the \ippdbcolumn{photcode} for which
the average lensing (and radial aperture) properties have been
calculated.

\paragraph{Image Tables} 

Measurements which are loaded into DVO may be associated with a
specific image (such as the measurements for a single chip from the
GPC1 camera) or they may not have such an association (such as
measurements from 2MASS, WISE, or externally supplied reference
measurements).  For data which is associated with an image, a subset
of the information about that image (e.g., from the header of the FITS
file) is used to populate a row in the DVO \ippdbtable{Image} table.
This table contains one row for each chip image known to DVO, with
information such as the filter (\ippdbcolumn{photcode}), the exposure
time, the airmass, the astrometric calibration terms, the photometric
zero point, etc.  For GPC1 and other mosaic cameras, an additional row
is defined to carry the projection and camera distortion elements of
the astrometry model.  As images are loaded into this table, they are
assigned an internal ID (a running sequence in the table), stored in
the field \ippdbcolumn{imageID}.  Images may also be assigned an ID
derived from the external source of the image (field
\ippdbcolumn{externID}): in the case of the GPC1 images, this ID is
defined by the processing mysql database and is guaranteed to be
unique within the processing system.  In the case of GPC1 exposures,
the external image ID is set to the database field
\ippdbtable{chipImfile}.\ippdbcolumn{chip_imfile_id}. A second field
(\ippdbcolumn{sourceID}) identifies which of the possible image-like
tables supplied this image, guaranteeing uniqueness of image IDs
across the different IPP stages.


\paragraph{Other Tables} 

Other tables are used to track information used by the calibration
system.  This includes the complete set of flat-field corrections
determined by the photometry calibration analysis and the astrometric
flat-field corrections determined by the astrometry calibration
analysis \citep[see][]{magnier2017.calibration}.

\subsubsection{Sky Partition}
\label{sec:SkyPartition}

\begin{figure*}[htbp]
  \begin{center}
 \includegraphics[width=\hsize,clip]{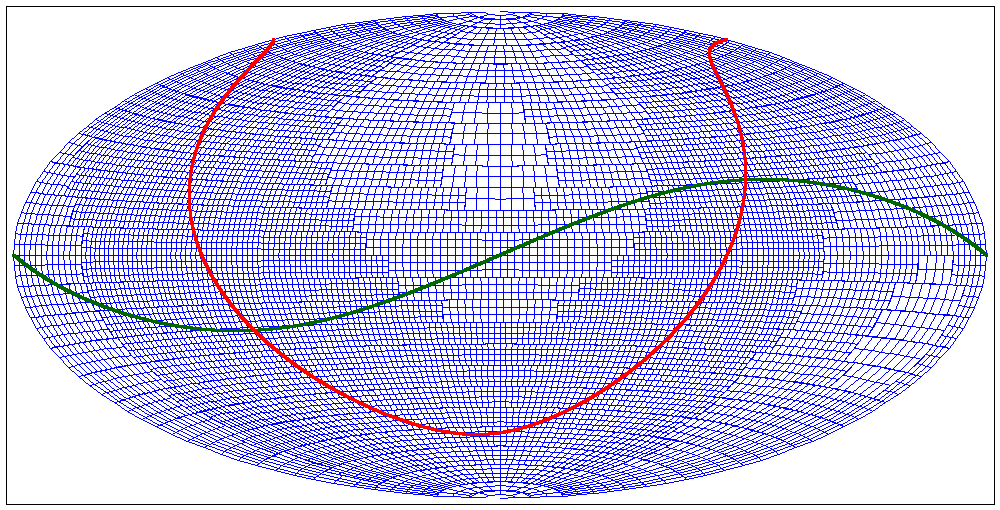}
  \caption{\label{fig:sky.partition} Level 3 sky paritioning.  The
    blue grid shows the outlines of the different regions assigned to
    separate tables in the sky partitioning scheme.  The Galactic
    plane is shown as a solid red line while the ecliptic is shown in
    green.  This organization of the sky duplicates that used by the
    HST Guide Star Catalog \citep{1988IAUS..133..239J}.  
 }
\end{center}
\end{figure*}

Tables within DVO containing information about astronomical objects
are partitioned on the basis of position in the sky: objects within a
region bounded by lines of constant RA,DEC are contained in a specific
file.  The boundaries and the associated partition names are stored in
one of the supporting tables, \ippdbtable{SkyTable}.  This table
contains the definitions of the boundaries for each sky region
(\ippdbcolumn{R_MIN}, \ippdbcolumn{R_MAX}, \ippdbcolumn{D_MIN},
\ippdbcolumn{D_MAX}), the name of the sky region, an ID
(\ippdbcolumn{INDEX}, equal to the sequence number of the region in
the table), and index entries to enable navigation within the table.
The regions are defined in a hierarchical sense, with a series of
levels each containing a finer mesh of regions covering the sky.

In the default used by the PV3 DVO, the partitioning scheme is based
on the one used by the Hubble Space Telescope Guide Star Catalog
files.  Level 0 is a single region covering the full sky.  Level 1
divides the sky in declination into bands 7.5\degree\ high, as defined
by the HST Guide Star Catalog
\citep[GSC][]{1988IAUS..133..239J,1990AJ.....99.2019L}.  Level 2
subdivides these declination bands in the RA direction, with spacing
related to the stellar density.  Level 3 divides these RA chunks into
4 - 8 smaller partitions (see Figure~\ref{fig:sky.partition}).  This
level exactly matches the HST GSC layout, and uses the same naming
convention to identify the partitions: \code{n0000/0000}, etc. Level 4
further divides these regions by a factor of 16.  In the
\ippdbtable{SkyTable}, a region at one level has a pointer to its
parent region (the one which contains it) and a sequence pointing to
its children (regions it contains).  The \ippdbtable{SkyTable} enables
fast lookups of the on-disk partitions which map to a specific
coordinate on the sky.  In general, a single DVO will have the full
sky represented with tables at a single level, although it is possible
for mixed levels to be used.  For the PV3 master database, the
partitioning is at Level 4, resulting in \approx 150,000 regions to
cover the full sky, of which \approx 110,000 are used for the PV3
$3\pi$ data.  The densest portions of the bulge contain at most
\approx 300,000 astronomical objects in the database files, with an
associated maximum of \approx 30 million measurements in these files.
With the compression scheme described below, the largest database
files are \approx 3GB, which can be loaded into memory in 30 seconds
on the processing machines that contain partition data.

The DVO software system allows the tables which are partitioned across
the sky to also be distributed across multiple computers, which we
call partition hosts.  A single file identifies these partition hosts
and the location of the database partition on the disks of that
machine.  The \ippdbtable{SkyTable} contains elements to define by ID
the partition host to which a set of tables has been assigned.
Operations which query the database, or perform other operations on
the database, are aware of the partitioning scheme and will launch
their operations as remote processes on the machines which contain the
data they need.  For example, a query for data from a small region
will launch sub-query operations on the machines which contain the
data overlapping the region of interest.  These remote query
operations will select the database information which matches the
query request (i.e., applying restrictions as defined) and return the
results to the master process.  The results from the various partition
hosts are then merged into a single result by the master process.
When the parallel partitioning for a DVO instance is defined, the
tables are randomly assigned to the partition hosts.  As a result,
queries which span more than a single partition are likely to spread
the I/O load across a large number of machines.  This parallelization
is critical to querying and manipulating the enormous database on a
reasonable timescale.

\subsubsection{Object and Measurement IDs}

Within the DVO system, certain integer fields are used to provide unique
identifiers for measurements and objects.  The original implementation
of DVO was limited to 32-bit integer fields, but since the maximum
number of objects and measurements was expected to be larger than
$2^{32}$, two 32-bit integer fields are joined together to make
sufficiently large IDs.

In the table of objects (\ippdbtable{Average}), the fields
\ippdbcolumn{objID} and \ippdbcolumn{catID} together form a unique
64-bit integer value to identify the objects.  The \ippdbcolumn{catID}
field is a sequence number for the sky partition table (the
`catalog') in which the object is contained, while \ippdbcolumn{objID}
is an incrementing sequence number within that sky partition
table.  As long as no sky partition tables contain more that
$2^{32}$ objects, these fields will not overflow.  These two fields
are included in the \ippdbtable{Measure}, \ippdbtable{GalPhot},
\ippdbtable{StarPar}, \ippdbtable{Lensing}, and \ippdbtable{LensObj}
tables to link the entries in those tables back their corresponding
object.  Note that \ippdbtable{SecFilt} does {\em not} contain these
ID fields; the rows in this table are maintained in the correct
sequence to match the \ippdbtable{Average} table entries.

The \ippdbtable{Measure} table, containing the detections of objects
from individual exposures or stack, or the (potentially
non-significant) measurements from a warp, uses the 32-bit integer
fields \ippdbcolumn{detID} and \ippdbcolumn{imageID} to uniquely
identify each entry.  The \ippdbcolumn{imageID} is the running
sequence number of the ``image'' (GPC1 OTA, stack, warp, or other
other source of the measurement) in which the object was measured.
The \ippdbcolumn{imageID} is a value internal to DVO, and is unique
across all types of images.  The \ippdbcolumn{detID} field is a 32-bit
integer giving the sequence number of the detection within the image.
For images processed by the IPP (e.g., using \ippprog{psphot}), the
\ippdbcolumn{detID} corresponds to the output field labeled as
\ippmisc{IPP_IDET} in those data products.  Since measurements from
the same image may be spread across multiple sky partition tables,
both \ippdbcolumn{detID} and \ippdbcolumn{imageID} much be used to
uniquely identify a detection within the database.  

In the \ippdbtable{Measure} table, the field \ippdbcolumn{averef}
specifies the row number in the \ippdbtable{Average} table of the
associated object.  The \ippdbtable{Measure} table may be unsorted, in
which case it is slow to find the measurements associated with a
specific object (a full table scan is required, referencing
\ippdbcolumn{objID}).  After the table is sorted and indexed, the
\ippdbcolumn{Measure} rows for a given object are grouped together.
In this case, the fields \ippdbtable{Average}.\ippdbcolumn{measureOffset} and
\ippdbcolumn{Average}.\ippdbcolumn{Nmeasure} define an index for the
code to jump to the list of measurements for a single object.  The
field \ippdbtable{Measure}.\ippdbcolumn{imageID} defines the link from
the measurement to the image which supplied the measurement.

DVO is also used to construct the unique object and detection IDs used
by the Published Science Products Subsystem (PSPS).  Within the PSPS,
the field named \ippdbcolumn{objID} in that database is used to
allows valid joins between tables to select the different kinds of
attributes of the same astronomical objects.  This 64-bit integer ID
is constructed based on the coordinates of the object, as described by
\cite[][]{flewelling2017}.  In short, the digits of the right
ascension and declination coordinates are used to define a single
64-bit integer with spatial resolution of roughly 3 milliarcseconds.
This values used by this field are generated by the DVO system and
stored in the \ippdbtable{Average} table in the field
\ippdbcolumn{extID}.  

Within the PSPS, the \ippdbtable{Detection} table carries an ID which
is unique for each measurement, equivalent to the DVO
\ippdbcolumn{det_id}, \ippdbcolumn{image_id} pair.  In this case, the
PSPS \ippdbcolumn{detectID} is constructed from the MJD of the
exposure, the number of the OTA (e.g., OTA64), and the detection
sequence within the image to form a single unique 64-bit integer value.
For detections from the stack images, the MJD is not unique, so a
different rubric is used to define IDs for those detections.  The
field \ippdbcolumn{XstackDetectID} (where '\ippdbcolumn{X}' is one of
g,r,i,z,y) is constructed from the GPC1 stack ID
(\ippdbtable{stackRun.stack_id}), the detection sequence within the
stack image, and the same value used to define \ippdbcolumn{sourceID}
above.  These two types of detection IDs are generated by the program
\ippprog{addstar} when the images and stacks are ingested into DVO.

\subsubsection{DVO Data Storage}

In the implementation of DVO used for the PV3 calibration analysis,
the database tables are stored on disk using binary FITS tables.  Each
type of database table is stored as a separate file, or a collection
of files for tables which are spatially partitioned.  The binary FITS
tables are compressed using the (to date) experimental FITS binary
table compression strategy outlined by \citet{2012arXiv1201.1340P}.  Table compression
is an option in DVO; for the PV3 database, the large data
volume (70TB compressed) drove the decision to compress the tables.

The FITS binary table compression scheme uses a strategy similar to
that used for FITS image compression
\citep[][]{1999ASPC..172..125W,2000ASPC..216..551P}.  The binary
tabular data is compressed and stored in the ``HEAP'' section of the
FITS table extension, with pointers to the compressed data stored in
the regular data section.  Each column in the FITS table is compressed
as one (or more) blocks.  The standard header keywords which describe
the data column format (e.g., TFORM1) are replaced with keywords which
describe the location and size of the compressed data in the HEAP
section; the information about the uncompressed data is moved to a
keyword with ``Z'' pre-pended (e.g., ZFORM1) and an additional field is
added to define the compression algorithm (e.g., ZCTYP1).  The column
names (e.g., TTYPE1) and units (e.g., TUNIT1) are retained in their
original form.

The compression algorithm can treat the entire column as a single
block of data, or it may be broken into a number of chunks, each
compressed in turn (this must be the same for all columns).
Additional header information is added to describe the block sizes and
information needed to describe the HEAP data section.  The compression
algorithms currently defined consist of the GZIP, RICE, PLIO, and
HCOMPRESS (REFS).  For GZIP, the compression algorithm may transpose
the byte order before compression: for floating point data of a
similar dynamic range, this choice may allow for better compression
as each byte in the 4 or 8 byte floating point value is more likely to
be similar to the same byte in other rows than to the other bytes of
the same row value.  This option is called \code{GZIP_2} in the FITS
standard, as opposed to the standard order, \code{GZIP_1}.  The DVO
system can be set to specify the compression options for each column
in the tables.  In practice, we have chosen a default in which
floating point numbers use \code{GZIP_2}, character strings use
\code{GZIP_1}, and integers use \code{RICE}.

\subsubsection{Addstar : DVO Ingest}
\label{sec:addstar}

Upon completion of the processing of each stage, the results of the
photometry analysis are stored in a large number of individual catalog
files as described in \cite{magnier2017.analysis}.  The data from
these files are loaded into a DVO database to define the astronomical
objects and to allow for calibration analysis.  The program which
loads the data into the DVO database is called \ippprog{addstar}, and
is associated with the the \ippstage{addstar} processing stage.  The
measurement catalogs generated by the \ippstage{camera},
\ippstage{skycal}, \ippstage{fullforce}, and \ippstage{diff} stages
are loaded into DVOs in this fashion, although not every measurement
in each catalog are included in the master DVO that is constructed.
For a particular re-processing version, a single master DVO is
constructed for the positive image stages (\ippstage{camera},
\ippstage{skycal}, \ippstage{fullforce}) and a separate one is
constructed for the difference image analysis stage results.

The construction of the master DVO is performed in a hierarchical
fashion.  The individual catalogs are added to a \ippmisc{minidvo},
which is simply a DVO database defined over some subset of possible
inputs.  These \ippmisc{minidvos} are then merged by stage into larger
databases to construct a single master DVO database.  In the process,
an intermediate master DVO for each stage is generated.  The
\ippprog{dvomerge} program is responsible for merging two DVO
databases together.  In the merge, astronomical objects are joined
together using essentially the same rules as those used to associated
detections into objects with one exception: the match radius may be
chosen to be a different size depending on the data source.  For
example, when WISE data is merged with PS1 data, as discussed below, a
match radius of 3 arcseconds is used due to the large beam-size of the
WISE telescope.

As of PV3, the process of merging \ippmisc{minidvos} is not highly
automated, requiring manual attention.  The generation of the
\ippmisc{minidvos} is automated and managed by the \ippstage{addstar}
stage.  Each catalog that is to be added to DVO has an entry created
in the \ippdbtable{addRun} database table.  This entry notes which
\ippdbcolumn{stage} is the source of the catalog, and links to the
appropriate database table with the \ippdbcolumn{stage_id} field.  As
some stages, such as the \ippstage{diff} stage, create more than a
single catalog, multiple entries with the \ippdbcolumn{stage_id} are
created, with the \ippdbcolumn{stage_extra1} field containing an index
to the individual components.  The catalog specified by the entry is
added to the target \ippmisc{minidvo} by the \ippprog{addstar}
program, updating the measurements in the appropriate DVO tables.
When this completes, an entry containing the statistics of the job is
added to the \ippdbtable{addProcessedExp} table.

After the master DVO is constructed containing the PS1 data, data from
other sources are also added to the database.  For the PV3 DVO
database, data was added from 2MASS, WISE, Gaia DR1, and Tycho.  These
external data sources are added by first generating a DVO database
containing just the particular data source, then using the same DVO
merging method to merge the external data DVO into the PS1 master.  

\subsection{Calibration Operations}
\label{sec:calibration}

Once the master DVO database has been constructed, high-quality
astrometric and photometric calibrations can be calculated.  The
details of the calibration analysis are discussed in
\cite{magnier2017.calibration}.  We present a brief summary here.

Astrometric calibration consists of measuring and correcting
systematic structures along with improved calibration of the
transformations from chip to focal plane coordinates based on relative
astrometry.  These steps are performed iteratively.  First, the
relative astrometry analysis generates an improved solution without
correction for systematic effects.  Next, systematic effects are
measured by querying the DVO database to determine the residual
astrometric error as a function of some parameters.  In the case of
the PV3 astrometry analysis, systematic errors have been determined as
a function of position in the camera (essentially an astrometric
flat-field correction), as a function of the brightness of the star
(the so-called Koppenh\"offer effect, see~\citealt{magnier2017.calibration}), and as
a function of airmass and color (differential chromatic refraction).
Once the systematic errors have been measured, they are applied back
to the measurements in the database.  Within the DVO
\ippdbtable{Measure} table, the different types of systematic effects
are included as separate offsets (in chip pixel coordinates) for each
measurement.  A single ``corrected'' version of the chip pixel
coordinates is stored in which the systematic offsets are combined
with the raw pixel coordinates for each measurement.  After the
systematic effects have been applied to the database, relative
astrometry is again performed this time using the corrected positions.

Photometric calibration consists of determination of zero points for
each exposure along with corrections for systematic effects.  In this
case, we rely on efforts of our external collaborators for the initial
zero point determination.  The team at CfA downloaded the per-exposure
catalog files (``smf files'') and determined the zero points of those
exposures which were believed to be obtained in photometric
conditions.  This process, called ``\"ubercal'', is described in
detail by \cite{2012ApJ...756..158S} for the first (PV1) version.  In
brief, photometric periods, with time-scales of a large fraction of a
night, are identified by a combination of automatic analysis and
manual inspection.  A single solution for all images in a given filter
is determined to minimize scatter for individual stars.  The free
parameters in this solution consist of a single zero point and airmass
slope for each photometric period along with a collection of
flat-field offsets for several large time range (``flat-field
seasons'').  For the PV3 \"ubercal analysis, the flat-field offsets
were determined on a $2\times2$ grid for each chip and 5 flat-field
seasons were identified.  The boundaries of the flat-field seasons
were determined by independent inspection of the residuals observed in
the Medium Deep fields.


After the \"ubercal analysis of the photometric periods is completed,
the determined zero points, airmass corrections, and flat-field terms
are transmitted back to the IfA IPP team.  These values are then
ingested into the master DVO database.  An initial relative photometry
analysis is performed to tie the images without \"ubercal zero points
to the \"ubercal system.  Zero points from the \"ubercal analysis are
not allowed to change, but zero points of the rest of the exposures
are determined to minimize the photometric scatter for bright stars.
These zero points are determined uniquely for each image.  After an
initial relative photometry analysis, the photometric residuals are
used to determine a systematic correction as function of position in
the camera.  This correction is equivalent to the flat-field
corrections determined as part of the \"ubercal analysis, but are much
higher spatial resolution ($40\times40$ corrections per chip) and are
determined for only the full time range of PV3.  This high-resolution
flat-field correction addresses photometric variations due to spatial
variations in the PSF due to the optics and low-level effects on the
chips \citep[see][]{magnier2017.calibration}.  After the systematic corrections
have been determined and applied back to the database, a final
relative photometry analysis pass is performed.

\subsection{Construction of the PSPS database}
\label{sec:ipp2psps}

The publicly-visible Pan-STARRS database is hosted by the Space
Telescope Sciences Institute through their Mikulski Archive for Space
Telescopes (MAST).  The underlying database at MAST is a copy of a
database generated at the IfA by the Published Science Products
Subsystem (PSPS).  The construction of the PSPS version of the PS1
database starts once the PS1 photometry and astrometry measurements
have been calibrated within the DVO system.  The construction takes
place in several stages, described in detail by \cite{flewelling2017}.
We summarize those steps here.

The first stage of constructing the PSPS database consists of the
generation of small files called ``batches'' which contain a complete
set of measurements for a small chunk of the database tables.  The
program which is responsible for the construction of these batches is
called \ippprog{ipptopsps}.  Several different types of batches are
generated, relating to the different types of tables in PSPS.  The
details of the batch construction depend on the batch type.  

One type of batch consists of measurements from the individual
exposures.  These batches are generated based on the output catalog
files generated at the \ippstage{camera} stage (``smf files'').  The
\ippprog{ipptopsps} program loads the complete set of measurements and
metadata from the smf catalog file, then queries the DVO database for
calibration parameters related to that smf file.  The batch is
constructed by applying the photometric calibrations to the raw flux
measurements in the smf file.

A second type of batch file consists of the measurements related to
the stack images.  Again, \ippprog{ipptopsps} starts with the output
catalog files, selects the appropriate calibration information from
the DVO, and applies the calibration data to the raw measurements in
the stack catalog files.

A third type of batch file consists of average properties of the
astronomical objects in the DVO database.  Unlike the other two batch
types, this operation is performed solely via queries to the DVO
database.  The complete set of average measurements for objects in a
single DVO spatial partition are loaded by \ippprog{ipptopsps} and
used to generate the batch file.  

As the batch files above are generated, the PSPS system can run in
parallel to ingest the measurements from these batch files.  PSPS
downloads in sequence the batch files as they are generated and
unpacks the data.  The data are then loaded into a small-scale version
of the PSPS database, using the full schema.  After a large chunk of
batches have been loaded, the resulting tables are then merged into
the master PSPS database.  After another large chunk of data has been
merged into the master PSPS database, a large-scale copy of the
database is made internally to provide a long-term backup and to aid
in error recovery.

Once the full PSPS database has been loaded, or a complete set of
batches for a given batch type, the entire database is copied to
STScI where it can then be made visible either to the Pan-STARRS
Science Consortium or to the wider public.

\section{Operations and Automation}
\label{sec:operations}

\subsection{Pantasks and Parallel Processing}
\label{sec:pantasks}

\subsubsection{Pantasks}

Sections~\ref{sec:stages} \& \ref{sec:postprocessing} describe the
analysis steps which take place in the Pan-STARRS data analysis
systems.  Individually, these steps appear as commands which could be
run by a user within the UNIX environment of the PS1 data system.  The
processing database (Section~\ref{sec:processing.database}) provides
the logical links to relate the results of one analysis stage to
another.  In order to make a complete system which can run
automatically, it is necessary to have a software system which can use the
contents of the processing database to generate the commands
corresponding to the analysis stages.  This system needs to (1)
regularly examine the database to find items from stages which are
ready to be processed, (2) have rules which define how to construct
the appropriate commands, (3) cause those commands to be executed
within the processing system, (4) monitor the active processing
jobs for completion, and (5) check on the results of those
commands and update the processing database as needed.  Within the
Pan-STARRS IPP, the top-level management of these operations is
performed by the program called \ippprog{pantasks}.  

The core capability of the \ippprog{pantasks} program is to take a
collection of ``tasks'' which describe the concept of a command which
might be run and to regularly generate new commands based on that
concept.  The ``tasks'' are defined using the \ippprog{opihi} scripting language
(also shared by DVO and other user-interactive programs within the
IPP).

\ippprog{Pantasks} repeatedly checks each task in an attempt to
generate a new command: we say \ippprog{pantasks} attempts to
``execute'' the task.  Tasks may specify the time between execution
attempts, with a 1 second default.

Each task must at a minimum define a command to generate.  Commands
may be static or dynamic.  For a task with a static command, the
command is explicitly defined in the task block (see code example in
Figure~\ref{fig:task_example}) and is identical each time the task is
executed.  A dynamic command is defined within a special block of the
task, called \code{task.exec}.  This block is a snippet of code (in the
\ippprog{opihi} language) which is run each time the task is executed.  The
\code{task.exec} code may refer to variables or other data structures
defined by the \ippprog{opihi} language within the \ippprog{pantasks}
environment.  Within a single \ippprog{pantasks} instance, all \ippprog{opihi}
variables and data structures have global context by default (\ie, all
are visible to all tasks).  Within the context of an \ippprog{opihi} macro
(equivalent of a function call), variables may be locally-scoped.
Other data structures (see below) are global and must be protected
with name space choices.

Within the \code{task.exec} macro, the command to be run is defined by
the script.  Once the \code{task.exec} macro exits successfully, the
defined command is then added to the list of jobs to be run within the
UNIX environment.  Jobs may be run in one of two ways: locally or via
the parallel processing system.  The task, or the \code{task.exec}
macro, uses the \code{host} command to define how to run the job.  If
the host is set to ``local'', then the job is run in the background by
\ippprog{pantasks} itself (using the C \code{execvp} function).
Otherwise, the job is sent to the parallel processing system to be run
on another machine within the cluster.  If the host is set to the
special value ``anyhost'', then the parallel processing system is
allowed to choose the processing computer arbitrarily.  Any other
value is taken to be the DNS name of the computer on which this job
should run.  The host may (optionally) be required for the command, in
which case the parallel processing system must ensure that the job
only runs on the specifically-named computer.  Otherwise, the parallel
processing system may choose to redirect the command to another
computer using its own rules, e.g. to balance processing load across
the cluster.

When the \code{task.exec} macro is run, the code may choose (e.g.,
based on tests of some global variables) to exit the macro with an
error condition.  In this circumstance, no job is produced by the
task.  The task will be tried again the next time it is executed.
This feature allows for the user to set processing blocks which depend
on some external tests.  For example, some task may check external
network connectivity and set a variable based on the network status;
other tasks may then choose to wait until the network is available
before attempting to run.

Other task options exist to control the system behavior in detail.
These options may be dynamically reset by the \code{task.exec} macro.
Some options control the number of jobs, such as limiting the number
of currently-outstanding jobs for a given task, or limiting the total
number generated.  Other options can be used to control the time when
jobs of a certain task are allowed to run.  It is also possible to
specify the UNIX ``nice'' level at which the job is
run when it is executed.  Finally, individual tasks may be disabled
while the system is still running.

\subsubsection{pcontrol}

Jobs which are generated by \ippprog{pantasks} may be run locally on
the machine running \ippprog{pantasks} or they may be distributed
across many machines in the computing cluster.  The parallel
processing system used by \ippprog{pantasks} is an independent
software system called \ippprog{pcontrol}\footnote{Alternatives are
  possible: e.g., {\tt condor} has been experimentally integrated with
  \ippprog{pantasks} for tests}.

This program is based on the same \ippprog{opihi} shell language used
by \ippprog{pantasks}.  The two programs communicate via a shared set
of pipes: \ippprog{pantasks} sends commands to the standard input of
the \ippprog{pcontrol}, and accepts back responses on the standard
output and standard error.  

\ippprog{pcontrol} maintains a list of jobs (commands to be run) and a
list of hosts (computers on which a job could be run).  Jobs may have
one of several states: pending (ready to run), running (jobs which are
running), exit (job has completed), busy (job is being checked by
\ippprog{pcontrol}), crash (job has exited with a signal, normally
\code{segv}).

Similarly, the hosts may also have one of several states: off, down,
busy, idle, etc.  A single host can accept a single job at a time.
Multiple host instances corresponding to the same machine may be
specified allowing a single computer to run more than one simultaneous
job.  

During operation, \ippprog{pcontrol} accepts new jobs from \ippprog{pantasks} and adds
them to the list of jobs to execute.  It also accepts from \ippprog{pantasks}
the names of computers on which it is allowed to run those jobs.

\subsubsection{pclient}

When \ippprog{pcontrol} is provided with the name of a computer, it will attempt
to make an connection to that machine via ssh.  When a
connection is made, the remote shell is used to run a special
interface program call \ippprog{pclient}.  This program accepts
command lines from \ippprog{pcontrol} and is responsible for executing the
individual commands in the local shell environment.  A single ssh
connection to a remote host keeps a single \ippprog{pclient} shell running for a
somewhat arbitrarly long time, executing many shell commands as needed.
This architecture avoids wasting overhead making the ssh connection to
the remote machine each time a command is executed, allowing for rapid
execution of many commands.  As a result, a single job within the IPP
architecture is allowed to be very light and short running if needed.

After \ippprog{pcontrol} sends a job (commands) to a specific \ippprog{pclient}, it checks
back occasionally to see if the command has been run and executed.  If
it has finished, then \ippprog{pcontrol} will query for the exit status, the
standard output and standard error streams from the command.  (where
do these go, back to \ippprog{pantasks}?), with the results associated with the
job statistics.  At that point, the \ippprog{pclient} on the remote machine is
ready to accept a new job from \ippprog{pcontrol}.  If any jobs are pending in
the list of jobs known to \ippprog{pcontrol}, it will send those jobs to any
machines which are idle.

While \ippprog{pcontrol} interacts with the many remote machines, it
occasionally interacts with \ippprog{pantasks} to report the results from the
jobs it has been monitoring.  \ippprog{Pantasks} occasionally requests a list of
the completed jobs.  It then requests the status information for each
completed job, including the standard error and standard output.  As
\ippprog{pantasks} receives this completion information, the jobs are removed
from the list managed by \ippprog{pcontrol}.  Thus \ippprog{pcontrol} maintains at most a
modest list of jobs which are ``in flight'' , leaving all interpretation
work to \ippprog{pantasks}.

At the \ippprog{pantasks} level, the tasks define how \ippprog{pantasks} should use the
exit status and output products from each job.  For example, the
stderr and stdout may be specified to go to a file (with static name
or name dependent on the specific job).  The task may define different
behavior depending on the exit code from the job.  

The \ippprog{pantasks} program can be run as a stand-alone program
which presents an \ippprog{opihi} shell interface to the user when it is
started.  This mode is useful for testing as all errors are reported
back to the \ippprog{opihi} shell.  However, when the user exits the shell, the
\ippprog{pantasks} instance exits, shutting down \ippprog{pcontrol} and all remote client
connections.  In standard operations, \ippprog{pantasks} is run in a client
server mode.  The server runs continuously in the background and
multiple users may connect via the \ippprog{pantasks_client} program.
Users can the send commands to the server to load scripts, add
parallel hosts, check status, and start or stop the \ippprog{pantasks}
operations. 

\begin{figure}
\begin{center}
\begin{verbatim}
task       example.static.task
  host     local
  command  ls /tmp
  periods  -exec 5.0
  npending 1
  stdout   /data/local/example.task.output
  stderr   /data/local/example.task.errors
end  
\end{verbatim}
\caption{\label{fig:task_example} Example of a simple static
  task in the opihi-based scripting language used by pantasks.  In
  this example, pantasks would run a single instance of the command
  ({\tt ls /tmp}) every 5 seconds, sending the stdout and stderr to
  the listed files. }
  \end{center}
\end{figure}


\subsubsection{Pantasks scripts: ippTasks}

\ippprog{Pantasks} provides an environment in which commands can be
generated and extensive parallel processing managed.  The details of
how to implement the different stages of IPP processing are captured
in a collection of scripts written for \ippprog{pantasks} in the
\ippprog{opihi} language.  In general, each stage is defined by an
associated script collected together under the \ippmisc{ippTasks}
collection.  While each script has its own details, there are a number
of common elements.

Most stages consist of two related tasks: a \ippmisc{load} task, which
is responsible to querying the processing database to identify entries
to be processed, and a \ippmisc{run} task, which is responsible for
managing the processing of the individual entries.

The \ippmisc{load} task for a particular stage generates
\ippmisc{load} jobs which query the processing database via a
dedicated database interface program (see the discussion of
\ippmisc{ippTool} in section \ref{sec:ipptools} below) for a list of
processing stage entries that are waiting to be run.  The
\ippmisc{load} jobs are executed on the host running the
\ippprog{pantasks} server.  Only one of each type of \ippmisc{load}
job is permitted to run simultaneously, preventing race conditions.

The results from the database query job are stored in an \ippprog{opihi} data
structure called a \ippmisc{book} within the \ippprog{pantasks}
environment.  Each row in the result set is saved to a separate entry
within the \ippmisc{book}.  These \ippmisc{books} are a hierarchical
associative array indexing the entries (\ippmisc{pages} to continue
the analogy) to be accessed via a particular key.  Keys for most
stages are a combination of the stage id and an identifier for the
individual component for the job that will be executed.  For a given
row in the result set, each column in the row is stored as a separate
line on the \ippmisc{page}, identified by the database column name.  An
additional line, the \ippdbcolumn{pantasksState}, is added so \ippprog{pantasks}
can manage the processing of the job which will be generated by this
page.  When the page is first generated, the
\ippdbcolumn{pantasksState} is set to \ippmisc{INIT}, indicating that
this \ippmisc{page} is a new addition to the \ippmisc{book}.  Once all
new \ippmisc{pages} have been added, the task then scans the
\ippmisc{book} for any pages with \ippdbcolumn{pantasksState} set to
\ippmisc{DONE}, and removes them from the book, as these represent
jobs that have finished.


The associated \ippmisc{run} task generates jobs constructed from the
collection of pages in the book.  The task examines the book and
selects the first available page with \ippdbcolumn{pantasksState}
of \ippmisc{INIT}.  The task uses the information in the page to
construct the appropriate command-line (e.g., lines in the page may
include input file names and output file names for the specific item
in the database).  The resulting command becomes a job in the \ippprog{pantasks}
collection of jobs.  Most IPP analysis stages specify that the jobs
are then sent to \ippprog{pcontrol} for parallel process.  Before task generates
the job, the \ippdbcolumn{pantasksState} is set to \ippmisc{RUN} so a
future execution of the task will not attempt to re-run this specific job.

Upon completion of the job, it is necessary to update the processing
database with the results, specifically indicating in the database
that the job has completed and if was successful.  Within the IPP,
this responsibility is left to the program which ran the analysis.
IPP analysis steps normally consist of two main elements: a C-language
program to do the data analysis work and a supporting Perl script
which performs the database update upon completion.  Upon completion,
the \ippprog{pantasks} \ippmisc{RUN} task is responsible for updating the
status within the book, but not within the processing database.  This
split keeps the interactions at the \ippprog{pantasks} level relatively light,
leaving the overhead of the database interaction within the job
running on one of the computing machines in the cluster.

In addition to these tasks, most stages have a \ippmisc{revert} task
paired with the \ippmisc{run} task.  These tasks run infrequently and
generate jobs which perform an operation on the processing database to
clear jobs which have failed with one of the ephemeral failure modes
(see the discussion in Section~\ref{sec:processing.database}).  This
step allows these failures to be cleared from the system, allowing
those jobs to be scheduled again.  

Similarly, some stages have \ippmisc{advance} tasks that update the
primary table to indicate that all of its components are complete.
For many of the early stages of the pipeline (the \ippstage{chip}
through \ippstage{warp} stages), this \ippmisc{advance} task also adds
an entry into the database table for the next stage of processing for
the exposure being considered.  This step allows the data to process
automatically from stage to stage without intervention.

The IPP processing database is used to manage all versions of an
analysis for all analysis stages.  In addition to the regular
processing of the nightly data products, there may be large-scale
re-processing analysis tasks or tests of various kinds.  It may be
necessary for a test analysis of a particular item to use a different
version of the processing software from the regular nightly analysis
(for example, when testing a new algorithm for release).  A mechanism
is needed to manage these different processing attempts of the same
items.  With the IPP, this is accomplished with an extra field,
\ippmisc{label}, for each processing stage.  Within the
\ippmisc{load}, \ippmisc{revert}, and \ippmisc{advance} tasks
discussed above, the query to the processing database for new items is
restricted to a set of user-defined labels.  A given instance of
\ippprog{pantasks} will be supplied a set of labels which are then applied to
all tasks managed by that \ippprog{pantasks}.  For example, the \ippprog{pantasks} which
manages the nightly processing of the basic science analysis stages
(\ippstage{chip} - \ippstage{warp}, \ippstage{stack}, \ippstage{diff}) is supplied with several labels which
correspond to the different kinds of observations being performed.  In
this way, the analysis of the nightly observations is kept separate
from other processing attempts.

\subsection{Stage automation}
\label{sec:automation}

Beyond of the basic sequence of \ippstage{chip} to \ippstage{warp},
there is no single natural ``next step''.  For example: a stack can be
generated with any number of input warps; a difference image can be
generated between a warp and any one of many other warps or stacks.
Without a single sequence, more complex and sophisticated decisions
much be made.

For nightly processing of data obtained at the summit, this is handled
by a set of ``nightly science'' tasks and an associated
\ippmisc{ippScript}.  These scripts have a well-defined and restricted
set of goals: to ensure that difference images are generated for each
exposure (either by pairing together warps or pairs warps with
pre-defined stacks), that nightly stacks are generated for MD fields,
and that the stacks are also differenced against an appropriate
reference.  

Pairing warps together is simplified by the observing strategy in
which the same pointing is observed multiple times in a night.  By
limiting to warp-warp pairs from the same pointing, the problem is
significantly reduced from the arbitrary case.  

Queuing the diffs is done by first examining the set of all
exposures that have been taken at the summit on the current night of
observing, and querying information from each stage up through
\ippstage{warp} stage.  These exposures are grouped by
\ippdbcolumn{filter} and \ippdbcolumn{object}, which is a unique
identifier for each telescope pointing on the sky.  Exposures in each
group are then sorted by increasing observation date
(\ippdbcolumn{dateobs}).  The database results for each stage
(\ippstage{chip}-\ippstage{warp}) are checked to ensure that the selected exposures have
been successfully processed for all stages through \ippstage{warp}.
Exposure groups are ignored until all exposures have either been
processed through warp or have failed with a bad quality, meaning the
exposure (or portion) cannot be processed.  Failed exposures are
rejected.  The remaining exposures are then paired sequentially, with
the final exposure ignored in the case of an odd number of accepted
exposures.  Exposures paired in this way are sent to the
ippstage{diff} analysis stage.

Once observations have been completed for the night (signaled by the
end-of-night dark exposures that are taken each morning after the
telescope closes), and the script has generated all \ippstage{diff}
pairs that can be made with the above rules, a second pass is
performed, this time with the exposures in each group sorted by
decreasing observation date.  This change in ordering allows exposures
that were excluded due to an odd number of exposures to be paired with
the exposure closest in time (with the exposure that was previously
first ignored).  Exposure pairs in which at least one exposure does
not have a pre-existing difference image are queued for difference
image analysis.

The nightly stacks are queued based on checking that a minimum number
of complete \ippstage{warp} entries exist for each filter and field.
For the nightly MD processing, this minimum number was set to 8
exposures, as this is the number of exposures taken for each field.
Once this number was reached, no more exposures are expected, so
\ippstage{stack} database entries can be queued from the
\ippstage{warp} entries.  Again, failures and weather can reduce the
number of usable exposures.  If no stack could be made for a given MD
field with the minimum number of inputs by the time of the
end-of-night darks, stacks are generated using whatever
exposures are available.

The automatic nightly processing ensures that data is processed as
soon as it is downloaded from the summit, reducing the lag between an
observation and the reduced data. 

The other processing task that requires automation is the reprocessing
of the entire $3\pi$ survey, as the size of the dataset make it
challenging to do manually.  To manage this, the ``large area
processing'' (LAP) task and script are used.  The first stage of this
processing is generating an entry in the \ippdbtable{lapSequence}
table defining a new reprocessing.  After this, individual
\ippdbtable{lapRun} entries can be queued that define a
\ippdbcolumn{filter} and a \ippdbcolumn{projection_cell} to be
considered.  These projection cells match the tangent plane centers
used for the warp tessellation.  A \ippdbcolumn{projection_cell} is a
unit of sky defined to be a square four degrees on each side which has
a single tangent plane projection \citep[][see]{waters2017}.
Once this
entry is defined, it is populated with all exposures (stored in the
\ippdbtable{lapExp} table in the database) that are located
within 5 degrees of the center of the projection cell included.  This
radius ensures that any exposure that overlaps the projection cell
will be included.  Once the exposures have been added, the other
exposures within the same sequence are checked to see if a
\ippstage{chip} stage entry has been generated, and if so, the
\ippdbcolumn{chip_id} for that entry is saved into the
\ippdbtable{lapExp} as well.  This linkage ensures that each exposure
is only processed once.  If no entry is found, a new \ippstage{chip}
entry is queued for processing.  The task periodically checks the
status of the exposures in each \ippdbtable{lapRun} entry, and if they
have all completed the \ippstage{warp} stage, then a \ippstage{stack}
is queued for each skycell contained within the
\ippdbcolumn{projection_cell}.

\subsection{Nebulous}
\label{sec:nebulous}

\subsubsection{Motivation and Concept}

A major concern recognized early in the Pan-STARRS project is the
challenge of storing and managing the large volume of data that is
generated by the GPC1 camera.  The \ippprog{Nebulous} system was
designed to aid in this process.  \ippprog{Nebulous} is not a file
system per-se, but only a method of tracking the locations of files
within the file system, and of tracking duplicate copies of the same
file.  The core of \ippprog{Nebulous} is a mysql database which tracks
``storage objects'', the equivalent concept of a file within the
system.  Each storage object may be associated with a number of copies
of the actual files on the disks in the storage system (called
``instances''), which are also recorded by the database.  In the IPP
cluster, the file instances are stored on a collection of computers
with substantial disk partitions shared via NFS.

\ippprog{Nebulous} also explicitly tracks the different computers on
which the file instances are stored.  This allows the system to expose
files to the user only on machines which are currently active in
\ippprog{Nebulous}.  If, for example, a storage computer crashes or
needs to be taken offline, the machine can be made unavailable in
\ippprog{Nebulous}, in which case only instances on other machines
will be supplied to users.

This localization is also useful for allowing the IPP processing to
target processing to computers based on the location of the data.  For
example, all raw images from a specific chip in the camera could be
stored on a specific computer (for at least one of the instances).
All of the analysis stages which interact with that chip could then be
preferentially targeted to be run on that computer.  The localization
in \ippprog{Nebulous} and the host targeted processing in
\ippprog{pantasks} can therefore work together to encourage processing
to require only local disk access, reducing the I/O local on the
network infrastructure.  In the early stages of the Pan-STARRS
project, this was important because network bandwidth was an expensive
resource.  In practice, the as-built IPP has had sufficient network
bandwidth that this targetting was not completely required.  In
practice, due to the timing of hardware acquisition, occasional
hardware failures, and other organizational details, targeted
processing has only been used to a moderate degree within the
Pan-STARRS cluster.

\subsubsection{Implementation Details}

The user interfaces to Nebulous consist of command-line programs as
well as APIs in both C and Perl.  The basic user commands to interact
with Nebulous are to 1) query the database for an existing storage
object, and find a valid file instance associated with that object; 2)
create a new storage object, which instantiates an empty file that can
be opened for writing; 3) replicate an existing storage object to
create more file instances; 4) cull a single file instance of storage
object from the cluster; and 5) remove a storage object, and ensure
that all file instances are removed.  The filehandles returned for
newly created instances can then be opened for reading and writing
data to that instance.


For the Nebulous users, the identifier of a storage object is a unique
string with the form of a UNIX file path: e.g. a/b/c/file.  When a
program creates a new file in \ippprog{Nebulous}, it supplies a URI of
the form \code{neb://HOST.VOL/PATH/FILE}.  The HOST and VOL(ume)
specifiers are optional, allowing a file to be created on a specific
computer (HOST) and disk (VOL).  The path and filename portions become
the identifier and are recorded in the \ippmisc{storage_object} table
in the \ippmisc{ext_id} field.  A storage object entry is then created
in the database for this id, and an instance of the file created on
the specified node.  If the host is unspecified, or if the specified
volume is full, then a host is chosen at random from available nodes.

Files are stored on specific computers in a \ippprog{Nebulous}
directory or directories on that computer.  In the IPP system, the
top-level Nebulous directories are usually placed at the root of the
storage device as mounted on the machine, in a subdirectory named
\code{nebulous}.  Beneath the top-level directory are 256
subdirectories with names of the form 00 - ff (i.e., 2 digit
hexadecimal number).  Each subdirectory has 256 subdirectories with
the same naming scheme.  

The filename of an instance in Nebulous is deterministic and derived
from the \ippmisc{ext_id}: the \ippmisc{ext_id} is hashed using
the SHA-1 function, and the first four hexadecimal digits of this hash
are separated into two two-digit strings and used as the top and
second level directory location for the disk file.  The instance table
in the Nebulous database includes a unique auto-incrementing index to
provide a unique SQL ID for each instance.  Under the subdirectory
identified above, the disk file name is by appending the database
instance id with a string derived from the \code{ext_id}: forward
slash characters are replaced in the name with colons so the string
can represent a file in the UNIX filesystem.  For the example URI
above, this results in a file located on disk in a location like
\code{/data/HOST.VOL/nebulous/d5/d8/42.PATH:FILE}.
This file naming structure has the benefit of providing redundancy
between the filename on disk and the instance in the database.

Nebulous tracks additional information beyond just the storage objects
and the associated instances.  As mentioned above, the storage volumes
are tracked to provide a link between a top level nebulous directory
and the computer which contains that directory.  The locations and
mount points for the actual NFS storage are listed in the
\ippdbtable{volume} table.  This table contains columns indicating if
the volume should be used for reading (\ippdbcolumn{available}) and
writing (\ippdbcolumn{allocate}).  As noted above, Nebulous will not
return a file to the user if the storage volume is marked as not
\ippdbcolumn{available}.  If a storage volume is marked as not to be
\ippdbcolumn{allocate}ed, then new storage objects will not generate
instances on that volume, but existing instances may be supplied to
the user.  

Another column, \ippdbcolumn{xattr}, is used to control the behavior
of this volume, with specific values used to denote desired behavior.
For instance, the volume may be marked to be used only for backup, in
which case it will not be used to store an instance by default, but
will be used preferentially if an instance noted as a backup when it
is generated.  Alternatively, a volume may be marked as permanently
unavailable, and should be ignored in most contexts.  This latter
option allows the system to retain the memory of hardware which has
been retired (and potentially to retain information about instances
which were previously on such machines). 

In addition to the static table describing the volumes, a second
dynamically-generated table, \ippdbtable{mountedvol}, lists those
volumes that are currently visible and accessible from the
\ippprog{Nebulous} database server.  This table also lists the total
and currently available disk space on each volume, allowing the
\ippprog{Nebulous} load balancing routines to prioritize those volumes
with large unused disk space before filling the volumes with only
small amounts remaining.  This table is updated every ten to twenty
minutes, after a scan of each of the volumes listed in the
\ippdbtable{volume} table.

The \ippdbtable{cabinet} table organizes the individual volumes into
``cabinets,'' a concept loosely based on the physical arrangement of
the storage servers in the data center.  These cabinets are used to
prevent the replication of a storage object within a group of volumes
where all instances of the object could be taken off line by a single
failure.  Since servers within a given cabinet in the data center
share a common set of power delivery units (PDUs), it is important to
ensure physical distance between replicated copies to guarantee that a
temporary failure of one of the cabinet PDUs does not significantly
impact processing.

The nebulous user APIs do not interact directly with the nebulous
mysql database.  Instead, they interact with one of several computers
with an Apache web server.  Interactions with the Apache server are
performed using the Simple Object Access Protocol (SOAP) interface,
while the Apache servers interact directly with the Mysql database
server.  This architecture avoids the overhead of setting up and
tearing down the Mysql connection for each Nebulous command, instead
using only the low-latency SOAP communications.

The Nebulous database currently (2017 July) contains information about
5,560,533,654 file instances for 3,543,240,981 storage objects.  All
raw data, along with permanent products such as catalogs and the
current versions of full-sky stacks, are replicated to ensure at least
two copies exist in case of hardware failure.  Based on the most
recent database ID values (which are unique and never reused), this
corresponds to roughly half of all the storage objects and file
instances ever created, due to the transient nature of many pipeline
products.


\subsection{Datastore repositories}
\label{sec:datastore}

Transferring data between the IPP and other parts of the Pan-STARRS
system is generally accomplished via a ``datastore'', an http service
that exposes data in a common form.  One of the main datastores used
by the IPP is the one located at the summit.  This datastore exposes a
list of the exposures obtained since the start of the PS1 operations.
Requests to this server may restrict to the latest by time.  Each row
in the listing includes basic information about the exposure: an
exposure identifier \citep[e.g., o5432g0123o; see][for
  details]{chambers2017}, the date and time of the exposure, the
telescope commanded pointing, the filter and exposure time, and the
observation comment for that exposure.  The row also provides a link
to a listing of the chips associated with that exposure.  This listing
includes a link to the individual chip FITS files as well as an md5
checksum.  Systems which are allowed access may download the raw chip
FITS files via http requests to the provided links.


The IPP also uses datastores to provide access to its own data
products.  The detections identified in the \ippstage{diff} stage
images are organized by the \ippstage{publish} stage, which writes
output files containing those detections to a datastore that is
monitored by the Moving Object Processing System
\citep[][MOPS]{2013PASP..125..357D}, which analyses the detections to
identify asteroids.  Separate datastores are also used by the
\ippstage{distribution} stage to provide access to data products to
the Pan-STARRS Science Consortium members.  

\subsection{ippTools and ippScripts}
\label{sec:ipptools}


The IPP relies on a number of common libraries and programs to handle
various tasks that are shared between multiple stages of the
processing.  These subsystems are described in this section, to
provide an introduction to these essential components that underlie
the rest of the pipeline.

As shown above, the \ippprog{pantasks} tasks rely on \ippmisc{ippTools}
calls for database queries.  Each stage has an appropriate
\ippmisc{ippTool}, allowing the database interaction to be governed by a
fixed set of inserts and queries.  Isolating the database interaction
in this way adds a layer of validation before queries are executed,
and ensures that all database modifications are handled in a uniform
fashion.  

In addition to simple queries and updates to entries already in the
database, the various \ippmisc{ippTools} programs can also be used to
define new processing runs for a stage.  Again, using pre-defined
queries wrapped in a program allow the options to be parsed to ensure
that any new processing run definitions are appropriately restricted
in scope, reducing the chance that mistakes will fill the database
with many unwanted jobs.

Connecting the \ippprog{pantasks} parallel processing environment to
the actual IPP analysis programs are a series of \ippmisc{ippScripts}
written in Perl.  These scripts are what are actually executed by
\ippprog{pantasks}, with command line options provided based on the
database query performed there by the \ippmisc{load} task.  These
options are combined with configuration information stored in
\ippmisc{ippconfig} recipe files.  

The appropriate recipe is selected from the configuration information,
based on the source camera of the data to be processed, and optionally
modified by the \ippdbcolumn{reduction} field in the database.  These
optional \ippdbcolumn{reduction} entries provide a way to group a
non-standard set of processing options together across multiple
stages, by selecting a recipe that is not the default.

With the set of configuration options and database entries for the
data to be processed, the \ippmisc{ippScript} checks the input files
that will be used, and confirms that a valid copy of each is available
from the \ippprog{Nebulous} system.  For stages that have a large
number of inputs (such as the \ippstage{stack} stage, which requires
images, masks, variance maps, and detection catalogs from each of the
potentially large number of \ippstage{warp} stage inputs), the input
files are organized into temporary input list files, formatted in an
appropriate way for the analysis program that will process them.

The script also sets up an output logfile for this processing run,
ensuring that any status information from either the script itself or
the underlying analysis is stored on disk.  The majority of this
information is identical between calls to the script, but for rare
failures of the analysis programs, retaining this information allows
for such problems to be diagnosed and repaired.  

The command line for the main analysis program is constructed based on
the database values, the recipe options, and the input file names.
The analysis program is then executed, and any failure reported back
to the parent \ippprog{pantasks} process.  In the standard case of the
analysis completing successfully, the script checks that the expected
output products were generated, preventing hidden I/O errors from
being a problem with subsequent processing of those output products.

One output product that must exist is the \code{stats} file, which is
generated by the analysis program and contains statistics about the
processing, including such things as the image background level, the
fraction of masked pixels, and the version numbers of the analysis
program.  This stats file is then parsed by the
\ippprog{ppStatsFromMetadata} program, which uses the information
within to generate command options for the \ippmisc{ippTool} program
to ensure that these statistics are included in the database row that
is created in the secondary database table for the individual
component processed.

\subsection{psLib and psModules}
\label{sec:pslib}

Underlying all of the analysis programs are the \ippmisc{psLib} and
\ippmisc{psModules} C libraries.  The more fundamental \ippmisc{psLib}
library defines the internal data structures that are used (arrays of
arbitrary type, vectors, images, and hash tables among others), manage
data access (particularly for FITS images and tables), and organize
string and error handling in a uniform fashion.  This library also
contains fundamental math operations, covering vector statistics,
matrix operations, and function minimization.  Common image operations
such as binning, interpolation, and convolution are also provided, as
well as the methods to to write JPEG versions of the data for
visualizations.  Finally, general coordinate transformations are
provided between planes and projections of spheres.

The functions provided by \ippmisc{psModules} have more focused scopes
that are nevertheless still shared between multiple programs.  The
isolation of source objects is included, providing the organization of
detections that is used in the \ippprog{psphot} photometry analysis
\citep{magnier2017.analysis}.  The PSF matching required for \ippstage{stack}
and \ippstage{diff} stage image combinations is as well.  The
unification of configuration options between config files on disk and
the options specified on the command line is handled by
\ippmisc{psModules} functions, as is the construction of data
structures in memory to represent the astronomical camera based on the
header information in the input file.  The functions to generate and
apply the detrend corrections to the data are also provided by this
library.

\section{IPP Hardware Systems}
\label{sec:hardware}

\subsection{Kihei Processing Cluster} 
\label{sec:kihei}

The majority of all Pan-STARRS processing has been performed on the
dedicated IPP cluster, located in Kihei on Maui.  This cluster was
originally located at the Maui High Performance Computing Center
(MHPCC), a United States Air Force research laboratory center managed
by the University of Hawaii.  This site was chosen based on the
original development funding provided by the Air Force Research Labs
\citep[see][for more details]{chambers2017}.  Once the Air Force
funding stopped being a significant driver for Pan-STARRS, the cluster was
physically moved from the MHPCC to a similar nearby computing center
located at the Maui Research and Technology Center.

The computing cluster is comprised of three main types of computers,
with a variety of individual specifications due to the cluster being
assembled from multiple generations of purchases.  The data storage
nodes contain several petabytes of storage space that are used to
store both the raw exposure data downloaded from the telescope as well
as processed data products.  These nodes are also used to do
processing, and have jobs targeted to them in an effort to reduce the
network I/O demands (see~\ref{sec:chip} for more on this process).

These storage nodes are not fully capable of completing all processing
on the short timescale necessary for each night's worth of data.  To
increase the processing capability, we have 212 ``compute'' nodes that
have small amounts of local storage, but are able to provide
additional processing power.  In addition to the direct processing of
image data, these nodes are also used to manage the \ippprog{Nebulous}
file interface, as well as controlling the job scheduling for the
processing.

The final type of computer in the cluster are the database servers.
These computers have large memory capacity and high-speed disk access
(originally fast spindle spinning disks, now migrated to SSDs) are
used to store and manage both the IPP gpc1 and \ippprog{Nebulous}
databases.  In addition to the main master servers, we have duplicate
servers used as database replicants, which allow for quick switching
from the main to backup servers in case of a hardware issue that
cannot be resolved immediately.

\subsection{Los Alamos National Labs} 
\label{sec:LANL}

In order to increase the processing rate for the $3\pi$ PV3 data, we
partnered with Los Alamos National Lab to gain access to the Mustang
supercomputer.  The supercomputer is comprised of 3088 processing
nodes, each with 12 cores and 64GB of RAM.  The processing nodes do
not have significant local disks, but are connected to multiple
petabyte scale scratch disks.  Job management is controlled by the
Moab HPC system\footnote{\url{http://www.adaptivecomputing.com/}},
which schedules resource requests among users, allocating processing
nodes to satisfy jobs, and terminating those jobs if they exceed their
scheduled time limit.

This system is part of the lab's ``Turquoise'' network, allowing it to
be used for research projects that do not handle sensitive data.  It
is, however, subject to stricter access controls than are in place at
the main IPP processing cluster.  Login sessions are terminated after
12 hours, requiring new sessions to be initiated regularly.  Network
access is also filtered, with only SSH connections allowed between the
IPP cluster and Los Alamos.  This restriction removes the ability for
the processing to contact the IPP processing database directly.

To work around this, additional steps were needed to ensure efficient
use of the computing resources.  A periodic poll of outstanding tasks
was done on the IPP cluster, using the information stored in the
database, and those tasks assigned to a processing bundle.  Each
component task in the bundle was then checked to identify the set of
input files needed to complete the task, the commands necessary to
complete the task, and the set of output files that should be
generated if the task completed successfully.  Once this information
had been generated for all tasks, the component lists were merged, and
the Moab job control file was constructed.

The control file contains the resource requests for the job, as well
as the commands to be executed to complete it.  The resource request
was calculated based on the number of tasks included in the job bundle
$N_\mathrm{tasks}$, and scaled by the expected execution time
($t_\mathrm{task}$) and computational intensity of the component tasks
($S_\mathrm{task}$).  For a given job bundle, an initial estimate of
the number of compute nodes needed is simply $\mathrm{nodes} =
S_\mathrm{task} * N_\mathrm{tasks} / 12$.  To ensure that jobs were
not prematurely terminated, we attempted to design the requested job
processing time to be 25\% longer than the expected time to complete
the component tasks.  Based on the initial node count, we calculated
the request time as $t_\mathrm{request} = \lfloor 1.25
\frac{\mathrm{nodes} * t_\mathrm{task}}{\mathrm{nodes}_\mathrm{max}}
\rfloor + 1$, where $\mathrm{nodes}_\mathrm{max}$ is the maximum
number of nodes that can be requested in a single job (1000 for
Mustang).  Table \ref{tab:SC_processing_parameters} contains the cost
values used for the various IPP processing stages.

\begin{table*}
\caption{\label{tab:SC_processing_parameters} Cost values for remote processing}
\begin{center}
\begin{tabular}{lcc}
\hline
\hline
{\bf IPP Stage} & {\bf $t_\mathrm{task}$ (s)} & {\bf $S_\mathrm{task}$} \\
\hline
  \ippstage{chip} & 150 & 2 \\
  \ippstage{camera} & 1700 & 2 \\
  \ippstage{warp} & 110 & 2 \\
  \ippstage{stack} & 1500 & 6 \\
  \ippstage{staticsky} & 7200 & 6 \\
  \ippstage{fullforce} & 300 & 2 \\
\hline
\end{tabular}
\end{center}
\end{table*}


Once the preparation for the job is complete, the input and output
file lists, the task list, and the job control file are transferred
via SCP to the Mustang cluster.  Local tasks are then initiated on the
user interface nodes to SCP the input files onto the scratch storage
disks if they do not already exist.  Once all the input files have
been copied, the job is submitted to Moab for scheduling.  The Moab
interface is periodically polled to determine the job status, and
after it has completed, the results are retrieved in a similar way.
Local tasks again SCP the output products, but to copy the results
back to the IPP cluster.

In addition to the standard output products, ``dbinfo'' files are
constructed as part of the job execution.  These files contain
database update commands to ensure that the IPP processing database
has the correct entries for the tasks that were remotely executed.
These commands are executed after confirming that all retrieved output
products are present.

Approximately half of the \ippstage{chip} through \ippstage{warp}
processing for the PV3 reduction was performed on Mustang, with
201,040 / 375,573 of the \ippstage{camera} stage products reduced
there.  Only processing through the \ippstage{stack} stage was
attempted, although with a smaller fraction of the total compared to
the \ippstage{camera} stage, with 290,257 / 998,886 being produced at
Los Alamos.  One reason for this decrease is that due to the memory
constraints on the Mustang processing nodes, we were unable to run
stacks with more than 25 inputs there.  Stacks with larger numbers of
inputs overflow the memory of the processing node, and as they do not
have disk space available for use as virtual memory, cause the machine
to hang until the job time limit is reached.  These stacks were
instead processed on the regular IPP cluster, where hosts with
sufficient memory were available.

\subsection{UH Cray Cluster} 
\label{sec:UHCray}

In December 2014, the University of Hawaii installed a 178-compute
node Cray supercomputer on the main Manoa campus.  As part of the
initial commissioning of this computer, Pan-STARRS was invited to use
this resource in February 2015, roughly corresponding with the
completion of the initial processing of the \ippstage{chip} through
\ippstage{stack} processing.  Although the number of nodes was much
smaller than that available on Mustang, the nodes were more robust,
with 20 cores and 128 GB of memory.  The scratch data storage was
somewhat smaller than at Los Alamos, with only a single 600 TB volume.
We had the unique ability to rapidly deploy to the UH Cray, using
almost all nodes for IPP processing as other users at the university
were designing code.  This rapid deployment was made possible by the
similarity of the Slurm\footnote{\url{https://slurm.schedmd.com/}}
scheduler and tools to those used by Moab (although the UH Cray has a
smaller $\mathrm{nodes}_\mathrm{max}$ of 10).

The UH Cray was used to do processing for the \ippstage{staticsky}
stage, running approximately half of that photometry (101,528 /
200,720).  We were also able to run part of the \ippstage{fullforce}
photometry there as well, although more had to be run on the IPP
cluster as other users started to utilize the system, with 168,685 /
994,890 runs processed there.

\section{Conclusion}

Since the Pan-STARRS\,1 telescope first came online in 2007, this
telescope has obtained 1.43 million exposures with GPC1, amounting to
a raw data volume of 4.32 petabytes.  The Pan-STARRS Image Processing
Pipeline has archived and processed these images on-the-fly to produce
discoveries of transient events and hazardous asteroids in real-time.
The IPP has been used to perform several re-processings of large
fractions of the science exposures to produce a well-calibrated data
release of the $3\pi$ Survey data.  To date, and including repeated
analysis, the IPP has processed 2.1 million exposures, detecting 900 billion
sources in those exposures (real and otherwise!).  The Pan-STARRS data
processing system represents a real example of astronomy data
processing on the very large scale.

\acknowledgments

The Pan-STARRS1 Surveys (PS1) have been made possible through
contributions of the Institute for Astronomy, the University of
Hawaii, the Pan-STARRS Project Office, the Max-Planck Society and its
participating institutes, the Max Planck Institute for Astronomy,
Heidelberg and the Max Planck Institute for Extraterrestrial Physics,
Garching, The Johns Hopkins University, Durham University, the
University of Edinburgh, Queen's University Belfast, the
Harvard-Smithsonian Center for Astrophysics, the Las Cumbres
Observatory Global Telescope Network Incorporated, the National
Central University of Taiwan, the Space Telescope Science Institute,
the National Aeronautics and Space Administration under Grant
No. NNX08AR22G issued through the Planetary Science Division of the
NASA Science Mission Directorate, the National Science Foundation
under Grant No. AST-1238877, the University of Maryland, and Eotvos
Lorand University (ELTE) and the Los Alamos National Laboratory.

\bibliographystyle{apj}



\end{document}